# Ultrafast Excited-State Energy Transfer in Phenylene Ethynylene Dendrimer: Quantum Dynamics with Tensor Network Method

*Sisi Liu[†], Jiawei Peng[†‡*], Peng Bao[§], Qiang Shi[§], Zhenggang Lan[†*]*

[†]SCNU Environmental Research Institute, Guangdong Provincial Key Laboratory of Chemical Pollution and Environmental Safety & MOE Key Laboratory of Environmental Theoretical Chemistry, South China Normal University, Guangzhou 510006, China and School of Environment, South China Normal University, Guangzhou 510006, China

[‡] School of Chemistry, South China Normal University, Guangzhou 510006, China

[§]Beijing National Laboratory for Molecular Sciences, State Key Laboratory for Structural Chemistry of Unstable and Stable Species, CAS Research/Education Center for Excellence in Molecular Sciences, Institute of Chemistry, Chinese Academy of Sciences, Zhongguancun, Beijing 100190, China and University of Chinese Academy of Sciences, Beijing 100049, China

**Abstract**

Photo-induced excited-state energy transfer (EET) processes play an important role in the solar energy conversions. The phenylene ethynylene (PE) dendrimers display great potential in improving the efficiency of solar cells, because of their excellent photo-harvesting and exciton-transport properties. In this work, we investigated the intramolecular EET dynamics in a dendrimer composed of two linear PE units (2-ring and 3-ring) using the full quantum dynamics based on the tensor network method. We first constructed a diabatic model Hamiltonian based on the electronic structure calculations. Using this diabatic vibronic coupling model, we tried to obtain the main features of the EET dynamics in terms of the several diabatic models with different numbers of vibrational modes (from 4 modes to 129 modes) and to explore the corresponding vibronic coupling interactions. The results show that the EET in the current PE dendrimer is an ultrafast process. Four modes with A' symmetry play dominant roles in the dynamics, other 86 modes with A' symmetry can damp the electronic coherence, and the modes of A" symmetry do not show the significant influence on the EET process. Overall, the first-order intrastate vibronic coupling terms show the dominant roles in the EET dynamics, while the second-order intrastate vibronic coupling terms give the visible impact here by damping the electronic coherence and slowing down the overall EET process. This work provides a valuable understanding of the physical insight in the EET dynamics of PE dendrimers.

# 1. Introduction

The dendrimeric molecules are a promising group of photovoltaic compounds for solar energy conversions, which are easy synthesized in a relatively well-controlled manner and exhibit superior properties of light harvesting and exciton transport.[1-7] Structurally, a typical dendrimer is composed of the linear conjugated branches with different lengths, such as the phenylene ethynylene (PE) units[8-12], which are joined by the meta-substitutions at the phenylene nodes. Such unique geometric characters can result in the electronic transition of the excited state being localized at the individual PE units and generate the local excited (LE) state.[13-15] In principle, the LE state at the short-length branch should have higher excitation energy than the one at the long-length unit. Therefore, once some higher LE states are prepared by the resonant radiation fields, the efficient multi-step and unidirectional excited state transfer (EET) processes should take place from the short branches to the long units.[16-19] The mechanism of the EET dynamics of the dendrimeric molecules received considerable research attentions over the last few decades, particularly the PE dendrimer.[20-26]

Experimentally, a variety of spectroscopic techniques were introduced to study the excited state processes of the dendrimeric molecules, the relevant observations confirm that the ultrafast EET process occurs from the short PE units to the long PE ones.[2, 20, 22-25, 27-31] Among these pioneer works, the absorption and fluorescence emission spectra were once used to investigate the energy transport dynamics in a class of dendrimers, and to explore the influence of the inclusion of the highly conjugated units. The experimental results indicate that the absorption and emission bands of the dendrimeric molecules are at the short and long wavelengths, respectively.[20] This indicates that an efficient funnel should exist to facilitate the intramolecular EET between different conjugated units. Besides, the femtosecond pump-probe transient absorption spectroscopy was also employed to detect the ultrafast excited-state dynamics in the nanostar dendrimer. The experimental observations show that the excited 2-ring units undergo the biexponential decays (3.0 ± 0.5 and 14 ± 2.5 ps), and the 3-ring and 4-ring branches have subpicosecond lifetimes.[27]

Theoretically, numerous efforts were also been made to study the static and dynamic properties of the dendrimeric molecules.[17, 18, 20, 23-25, 32-37] For instance, several groups, including Mukamel, Chernyak, Martínez, Bardeen, Yoshizawa and others co-workers studied the main photophysical features of various some dendrimers, including the molecular orbital interactions, electronic couplings and exciton localization.[14, 15, 30, 38-40] For instance, for the PE dendrimer composed of the 2-ring and 3-ring PE units, the work by Huang et al[36] once employed various electronic structure methods, including the SCS-ADC(2), HF/TDHF, DFT/TDDFT with different functionals (B3LYP, BH&HLYP, CAM-B3LYP), and DFT/MRCI, to perform the benchmark calculations, to obtain the excited-state energies, to characterize the electronic characters, and to clarify the dominant role of the stretching motion of the C≡C bond of the 2-ring in the EET process.

To get the deep understanding of the EET dynamics of PE dendrimers, Mukamel and co-workers used the reduced density matrix framework to study the ultrafast exciton motion and time-dependent spectra, and pointed out their relationship.[10, 14, 21, 38, 41] Besides, a number of studies have been done by the Roitberg, Fernandez-Alberti, Tretiak and Shalashilin groups (alone or by their collaboration).[15, 19, 24, 25, 31, 42-52] For instance, they performed a series of systemic studies on intramolecular EET process in the PE dendrimer, using different on-the-fly dynamics methods ranging from the surface hopping to multi-configurational Ehrenfest approaches at the semi-empirical AM1/CIS level. These works clearly demonstrate the bright excited states mainly involve the local excitation at different branches, and the efficient unidirectional intramolecular EET process from the shorter units to the longer ones is governed by the ultrafast nonadiabatic transition. Specifically, they pointed out that the stretching motion of the ethynylene triple bond in the 2-ring units plays an important role in the EET dynamics for the PE dendrimer composed of the 2-ring and 3-ring units. Recently, Desouter-Lecomte et al.[53, 54] developed a simplified but effective PE dendrimer model by retaining the bright electronic states only and considering the specific vibrational domain. On this basis, they studied the early dynamics of the interaction hierarchy, and analyzed the effect of the in-plane high frequency skeletal vibration mode involving triple bonds. In addition, to analyze experimental observations, several time-resolved spectroscopic studies were conducted.[21, 37, 55] Hu et al.[37] simulated the time-resolved polarization-sensitive transient-absorption pump−probe spectra using the on-the-fly surface-hopping nonadiabatic dynamics, which demonstrates the ultrafast EET process from the 2-ring to 3-ring units can be monitored directly by employing the pump and probe pulses with different polarizations.

Although many excellent theoretical studies above have been conducted on the excited-state dynamic of the PE dendrimers, it is still essential to address the critical roles of vibrionic coupling effects in the corresponding EET dynamics. However, the precise description of such vibronic coupling effects is not a trivial problem. On the one hand, it is well known that the mixed quantum classical methods suffer from their inherent defects, for instance the failure in the treatment of electronic coherence in the nonadiabatic dynamics. On the other hand, the quantum dissipative dynamics generally does not give the explicit treatments on many vibrational modes. Alternatively, the full quantum dynamics methods may provide the accurate description on the ultrafast EET dynamics. However, the computational cost of the full quantum dynamics is generally very high. In this sense, it is rather challenging to study the ultrafast EET dynamics with high dimensions. One effective method to address this challenging is the multiconfiguration time-dependent Hartree (MCTDH)[56-61] method and its multilayer extension (ML-MCTDH).[62-75] This method is widely used to understand the electron or energy transfers in the excited-state process of organic photovoltaic systems.[76-88]

It is noted recently that the tensor network methods have gained great attentions in the simulation of the nonadiabatic dynamics of complex systems due to their high accuracy and efficiency.[89-123] In mathematical formalization, the ML-MCTDH method also belongs to a kind of tree-structure tensor network.[67] Alternatively, one of the representative approaches in this family is the chain-like tensor network so-called tensor train algorithm, in which the wave functions and operators are expressed in terms of the matrix product format, i.e. the matrix product state (MPS) and matrix product operator (MPO), respectively. Then, the efficient and accurate time evolution algorithms can be used to propagate the full quantum dynamics of the multi-dimensional systems. In practice, the MPS framework sometimes may break the quantum exponential barrier, if the bond dimension grows only polynomially with the system size. Thus, the tensor network method is an ideal tool to investigate the EET process in the dendrimer.

In this work, we tried to study the ultrafast EET dynamics of the dendrimer. Here, the dendrimeric molecule composed of two linear PE units (2 rings and 3 rings) linked by a phenylene node with meta-substitution was chosen as the prototype to study the EET process. First, the ground and excited states of this system were calculated based on the *ab initio* method, including the frequencies, coordinates of the dimensionless normal modes and PESs. Then, the diabatic model Hamiltonian of the PE dendrimer was constructed by fitting, in which total 129 vibrational modes were included. Finally, the tensor network method based on the train structure was used to investigate the nonadiabatic behavior in the EET dynamics, and the vibronic coupling effects were further explored in terms of a series of reduced models. The dynamics results confirmed that the EET in the PE dendrimer is an ultrafast process, and the vibration modes with different symmetries played specific roles in the EET process. Particularly, the modes with A" symmetry had little influences on the EET process. Moreover, the second-order intrastate vibronic coupling terms in the model can damp the electronic coherence and slow down the overall EET process.

This article is organized as follows. The theoretical methods and computational details are introduced in section 2, including the construction of diabatic model Hamiltonian, the tensor network method, the electronic-structure calculations and dynamics simulations. The results and discusses of diabatic Hamiltonian and EET dynamics are given in section 3. The conclusion and outlook are given in section 4.

## 2. Theoretical Methods and Computational Details

### 2.1. Diabatic Hamiltonian

The diabatic Hamiltonian $\mathbf{H}^{dia}$ is written as follows:

$$\mathbf{H}^{dia} = \mathbf{T}_{nuc} + \mathbf{V}_{el}^{dia}, \tag{1}$$

where $\mathbf{T}_{nuc}$ represents the kinetic energy of nuclei, and $\mathbf{V}_{el}^{dia}$ denotes the electronic potential energy in the diabatic representation. For the PE dendrimer studied in this work, the excited-state process mainly involves the EET from the local excited state at the 2-ring unit (LE2) to the local excited state at the 3-ring unit (LE3), thus only these two LE states were considered here, and the vibronic coupling Hamiltonian in the diabatic representation can be further expressed as:

$$\mathbf{V}_{el}^{dia} = \begin{bmatrix} V_{11}^{dia} & V_{12}^{dia} \\ V_{21}^{dia} & V_{22}^{dia} \end{bmatrix}, \tag{2}$$

with

$$V_{11}^{dia} = V_{11}^{(0)} + \frac{1}{2}\sum_i^N \omega_i Q_i^2 + \sum_i^N \kappa_i^{(1)} Q_i + \sum_i^N \gamma_i^{(1)} Q_i^2, \tag{3}$$

$$V_{12}^{dia} = V_{21}^{dia} = \lambda, \tag{4}$$

$$V_{22}^{dia} = V_{22}^{(0)} + \frac{1}{2}\sum_i^N \omega_i Q_i^2 + \sum_i^N \kappa_i^{(2)} Q_i + \sum_i^N \gamma_i^{(2)} Q_i^2, \tag{5}$$

where $V_{11}^{(0)}$ and $V_{22}^{(0)}$ are the energies of two diabatic electronic states at the ground-state minimum, respectively. $N$ represents the number of dimensionless normal modes. $Q_i$ is the dimensionless normal coordinate of Mode $i$ with the associated frequency $\omega_i$. $\kappa_i^{(n)}(n=1,2)$ is the first-order intrastate linear coupling constant of the $i$-th mode on the $n$-th electronic state, which tunes the energy gap between two electronic states and leads to the crossing of potential energy surfaces; $\gamma_i^{(n)}(n=1,2)$ is the second-order intrastate coupling coefficient for Mode $i$ on the $n$-th electronic state. $\lambda$ is the diabatic coupling constant quantifying as the interaction between different diabatic electronic states.

The PE molecular studied in this work displays the point group of Cs symmetry, which processes two irreducible representations A' and A". Therefore, the symmetry selection rules can be used to identify the non-zero vibronic parameters in the diabatic Hamiltonian. As the electronic Hamiltonian is always totally symmetric, the linear electron-phonon coupling constant in the diagonal Hamiltonian elements may be nonzero only for the totally symmetric coordinate. For the second-order electron-phonon coupling constant, all modes should contribute to this term since the square of the normal coordinates should always give the A' symmetry.

## 2.2. Construction of Diabatic Model

The diabatic model was constructed by fitting to the *ab initio* results. For the current two-state system, the adiabatic-to-diabatic transformation is satisfied with the following form:

$$V_+^{adia} + V_-^{adia} = V_{11}^{(0)} + V_{22}^{(0)} + \sum_i^N \omega_i Q_i^2 + \sum_i^N \left(\kappa_i^{(1)} + \kappa_i^{(2)}\right) Q_i + \sum_i^N \left(\gamma_i^{(1)} + \gamma_i^{(2)}\right) Q_i^2, \quad (6)$$

$$\left(V_+^{adia} - V_-^{adia}\right)^2 = \left[\left(V_{11}^{(0)} - V_{22}^{(0)}\right) + \sum_i^N \left(\kappa_i^{(1)} - \kappa_i^{(2)}\right) Q_i + \sum_i^N \left(\gamma_i^{(1)} - \gamma_i^{(2)}\right) Q_i^2\right]^2 + 4\lambda^2. \quad (7)$$

Here, $V_\pm^{adia}$ are the adiabatic energies of two electronic states, which can be easily obtained in terms of the common quantum chemistry software at different levels. By performing fits to the values of $\left(V_+^{adia} + V_-^{adia}\right)$ and $\left(V_+^{adia} - V_-^{adia}\right)^2$, the diabatic Hamiltonian parameters could be obtained in principle. However, since many variables were involved in the fitting, this process was not a trivial problem. Thus, we adopted the following procedure in this work.

Firstly, according to the Eq. (7), the parameter $\lambda$ could be calculated as the minimum-energy gap of two excited states along different dimensionless normal coordinates. As two lowest adiabatic states mainly display the LE2 and LE3 characters, it was enough to choose them in the construction of the diabatic Hamiltonian model. Thus, we scanned the PES along different dimensionless normal coordinates to locate the geometry with the minimum adiabatic energy gap of two low-lying excited states. And the half of the energy gap at this geometry defines the parameter $\lambda$. After the scan of all dimensional coordinates, we chose the minimum $\lambda$ to define its final value.

Next, both the first-order linear and second-order non-linear vibronic coupling coefficients can be obtained by fitting. For each normal mode, the corresponding $V_{11}^{(0)}$, $V_{22}^{(0)}$, $\kappa_i^{(n)}$ $(n=1,2)$ and $\gamma_i^{(n)}$ $(n=1,2)$ were obtained in terms of fitting to $\left(V_+^{adia} + V_-^{adia}\right)$ and $\left(V_+^{adia} - V_-^{adia}\right)^2$ along individual normal coordinate. In this step, we fixed the frequencies $\{\omega_i\}$ and interstate electronic coupling constant $\lambda$. Since the obtained $V_{11}^{(0)}$ and $V_{22}^{(0)}$ had various values from the fitting results along different normal coordinates, we averaged all obtained values to define the final $V_{11}^{(0)}$ and $V_{22}^{(0)}$. Then, we fixed these available terms and re-fit the first-order and second-order vibronic coupling terms again to get the final multi-dimensional model.

To check the validity of diabatic model Hamiltonian, we recalculated the adiabatic potential energy surface along each normal coordinate and compared them with the *ab initio* results, in which the root mean square error (RMSE) between them was used to estimate the quality of the fitting results.

## 2.3. Tensor Network Method

The tensor network method is a very efficient dynamics approach, which can greatly alleviate the cure of dimensionality in quantum dynamics evolution. In this work, we use the tensor network based on the train structure to propagate the dynamics. The time-dependent wave function $|\Psi(t)\rangle$ of a molecular with $L$ sites can be expressed in terms of the d-dimensional local state spaces $\{\sigma_i\}$ as follow,

$$|\Psi\rangle = \sum_{\sigma_1,...,\sigma_L} \mathbf{C}_{\sigma_1...\sigma_L} |\sigma_1...\sigma_L\rangle$$

$$= \sum_{\sigma_1,...,\sigma_L} \sum_{\alpha_1,...,\alpha_L} \mathbf{M}^{\sigma_1}_{1,\alpha_1}...\mathbf{M}^{\sigma_i}_{\alpha_{i-1},\alpha_i}...\mathbf{M}^{\sigma_L}_{\alpha_{L-1},1} |\sigma_1...\sigma_L\rangle \qquad (8)$$

$$= \sum_{\sigma_1,...,\sigma_L} \mathbf{M}^{\sigma_1}...\mathbf{M}^{\sigma_i}...\mathbf{M}^{\sigma_L} |\sigma_1...\sigma_L\rangle,$$

where the coefficient $\mathbf{C}_{\sigma_1...\sigma_L}$ is a $d^L$-dimension tensor and can be decomposed in the form of matrix product. $\mathbf{M}^{\sigma_i}$ represents a single matrix for the $i$-th site, which is a rank-3 tensor with two matrix indices ($\alpha_{i-1}$ and $\alpha_i$) and one physical index ($\sigma_i$). The maximum values of the matrix and physical indices are termed as the bond dimension (BD) and physical bond (PB), respectively.

Applying the singular value decomposition (SVD) to the tensor, the wavefunction can be rebuilt as a left-canonical (right-canonical) MPS consisting only of left-normalized (right-normalized) tensor or the mixed-canonical MPS, which can greatly increase the numerical stability of algorithm and simplify contractions, namely

$$|\Psi\rangle = \sum_{\sigma_1...\sigma_L} \mathbf{L}^{\sigma_1}...\mathbf{L}^{\sigma_i}...\mathbf{L}^{\sigma_L} |\sigma_1...\sigma_L\rangle$$

$$= \sum_{\sigma_1...\sigma_L} \mathbf{R}^{\sigma_1}...\mathbf{R}^{\sigma_i}...\mathbf{R}^{\sigma_L} |\sigma_1...\sigma_L\rangle \qquad (9)$$

$$= \sum_{\sigma_1...\sigma_L} \mathbf{L}^{\sigma_1}...\mathbf{L}^{\sigma_{l-1}}\mathbf{S}^{\sigma_l}\mathbf{R}^{\sigma_{l+1}}...\mathbf{R}^{\sigma_L} |\sigma_1...\sigma_L\rangle.$$

Here, $\mathbf{S}^{\sigma_l}$ represents the correspond tensor on the site $l$, $\mathbf{L}^{\sigma_i}$ and $\mathbf{R}^{\sigma_i}$ are the left-canonical and right-canonical tensors, respectively.

Correspondingly, the operator can also be expressed as MPO on the same local state spaces with MPS,

$$\hat{O} = \sum_{\sigma_1,...,\sigma_L} \sum_{\sigma_1',...,\sigma_L'} \sum_{b_1,...,b_L} \mathbf{W}^{\sigma_1\sigma_1'}_{1,b_1}...\mathbf{W}^{\sigma_i\sigma_i'}_{b_{i-1},b_i}...\mathbf{W}^{\sigma_L\sigma_L'}_{b_{L-1},1} |\sigma_1...\sigma_L\rangle\langle\sigma_1'...\sigma_L'|$$

$$= \sum_{\sigma_1,...,\sigma_L} \sum_{\sigma_1',...,\sigma_L'} \mathbf{W}^{\sigma_1\sigma_1'}...\mathbf{W}^{\sigma_i\sigma_i'}...\mathbf{W}^{\sigma_L\sigma_L'} |\sigma_1...\sigma_L\rangle\langle\sigma_1'...\sigma_L'|, \qquad (10)$$

where $\mathbf{W}_{b_{i-1},b_i}^{\sigma_i,\sigma_i'}$ ($b_0 = b_L = 1$) is a rank-4 tensor with two matrix indices ($b_{i-1}$ and $b_i$) and two physical indices ($\sigma_i$ and $\sigma_i'$).

In this work, the time-dependent variational principle (TDVP) had been chosen to solve the time-dependent Schrödinger equation (TDSE) and perform the dynamics evolution, based on the representations of MPS and MPO. In the one-site TDVP algorithm, the evolution is constrained by a manifold of predefined MPS and obtained by a projection onto the tangent space. Inserting the projector $\hat{P}_{T_{|\Psi(t)\rangle}}$ onto TDSE, the equation becomes

$$i\hbar \frac{\partial}{\partial t}|\Psi(t)\rangle = \hat{P}_{T_{|\Psi(t)\rangle}} \hat{H} |\Psi(t)\rangle$$
$$= \sum_{i=1}^{L} \hat{P}_{i-1}^{L,|\Psi(t)\rangle} \otimes \hat{I}_i \otimes \hat{P}_{i+1}^{R,|\Psi(t)\rangle} \hat{H} |\Psi(t)\rangle - \sum_{i=1}^{L} \hat{P}_{i}^{L,|\Psi(t)\rangle} \otimes \hat{P}_{i+1}^{R,|\Psi(t)\rangle} \hat{H} |\Psi(t)\rangle, \quad (11)$$

where $\hat{P}_{i-1}^{L,|\Psi(t)\rangle}$ and $\hat{P}_{i+1}^{R,|\Psi(t)\rangle}$ are the left and right projector, respectively,

$$\hat{P}_{i-1;\bar{\sigma}_1,...,\bar{\sigma}_{i-1};\sigma_1,...,\sigma_{i-1}}^{L,|\Psi(t)\rangle} = \sum_{m_{i-1}} \overline{\Psi}_{i-1;m_{i-1}}^{L;\bar{\sigma}_1,...,\bar{\sigma}_{i-1}} \otimes \Psi_{i-1;m_{i-1}}^{L;\sigma_1,...,\sigma_{i-1}}, \quad (12)$$

$$\hat{P}_{i+1;\bar{\sigma}_{i+1},...,\bar{\sigma}_L;\sigma_{i+1},...,\sigma_L}^{R,|\Psi(t)\rangle} = \sum_{m_i} \overline{\Psi}_{i+1;m_i}^{R;\bar{\sigma}_{i+1},...,\bar{\sigma}_L} \otimes \Psi_{i+1;m_i}^{R;\sigma_{i+1},...,\sigma_L}. \quad (13)$$

In the propagation, the TDSE was approximated by solving each term individually and sequentially, then multiplying each individual equation above by the single-site map $\overline{\Psi}_{i-1}^{L} \otimes \overline{\Psi}_{i+1}^{R}$ or the center-bond map $\overline{\Psi}_{i}^{L} \otimes \overline{\Psi}_{i+1}^{R}$, respectively. Instead of having to work with the full MPS $|\Psi\rangle$, we worked with the effective single-site and effective center matrix tensors and associated local Schrödinger equations directly. The population and the electronic coherence are calculated as the trace of $\langle \phi_i|\Psi\rangle\langle\Psi|\phi_j\rangle$ over each vibronic degrees of freedom, *i.e.*, $\rho_{i,j}^{\text{ele}} = \text{Tr}\{\langle\phi_i|\Psi\rangle\langle\Psi|\phi_j\rangle\}$. When *i* is equal to *j*, $\rho_{i,j}^{\text{ele}}$ represents the population, otherwise, it represents the coherence.

The increase of matrix dimension is inevitable in the simulation, thus, the truncation is a necessary way to control the dimension and decrease the computational effort, which includes SVD compression and variational compression. In the tensor network method based on the train structure, the SVD compression is often used to control the bond dimension, which also has two ways. One is to truncate every matrix by constant from side to side, and the other is to truncate the total dimension making it smaller than a constant. Besides, the order of the sites is also important for the efficiency of the calculation. For the simple systems, we can set the sort naively. While for the complex systems, the chain mapping or its extended start-type approach would be a good choice.[94, 124, 125]

### 2.4. Computational Details

All electronic structure calculations were preformed using the Gaussian 16 package.[126] For the ground-state optimization and frequency analysis, we employed DFT at the ωB97XD level with the 6-31G* basis set (DFT/ωB97XD/6-31G*). This step generated the ground state equilibrium geometry $S_0^{min}$ and dimensionless normal coordinates. The excited-state PESs along the dimensionless normal coordinates were built at the time-dependent density functional theory (TDDFT) level using the ωB97XD functional and the 6-31G* basis set (TDDFT/ωB97XD/6-31G*).

During the construction of diabatic model, the least square method achieved in Gnuplot program was adopted.[127] Based on this method, the potential energy curve was fitted by minimizing the sum of squares of the errors of the data points to the fitted curve.

In the dynamic simulation, the initial electronic state was prepared by vertically putting the lowest vibrational level of the electronic ground state to the diabatic LE2 state. The whole population dynamics within 200 fs was simulated with a step size of 0.25 fs.

As analyzed in the below discussion, modes $v_{82}$, $v_{114}$, $v_{118}$ and $v_{120}$ play very important roles in the dynamics. Thus, to improve the efficiency in the tensor-train expansion of the total wavefunction, we put modes $v_{82}$ and $v_{114}$ in the left side of the whole chain, modes $v_{118}$ and $v_{120}$ in the right side of the whole chain, and the other modes in the middle. Here, PB and BD are important to the converged dynamics results, which were confirmed in terms of the testing. In the current work, all BDs were set as 250, the PBs of modes $v_{82}$, $v_{114}$, $v_{118}$ and $v_{120}$ were set as 250, and the PBs of other modes were set as 20.

The computational efforts in the dynamics simulations are given as follows. For the calculations on CPU E5-2678, the 4-mode dynamics needed 4 h, the 90-mode dynamics required 35 days, and the 129-mode dynamics needed 60 days. For the calculations on the GPU 3090, the 4-mode dynamics needed 4 h, the 90-mode dynamics required 20 days, and the 129-mode dynamics needed 30 days.

## 3. Results and Discussion

We investigated the ultrafast EET process from 2-ring unit to 3-ring one in the prototype PE dendrimers. To understand the EET process in depth and detail, the first step in this work was to perform the excited-state quantum chemistry calculations, which can obtain the basic electronic characteristics of both ground and excited states. Then, we constructed the diabatic vibronic model Hamiltonian. On this basis, we simulated the EET process by using the tensor network method.

### 3.1. Electronic Structure Calculation

The PE dendrimer studied in this work is composed of two linear PE units (2 rings and 3 rings), in which the adjacent benzene rings are connected with the C≡C bond. The equilibrium structure of the electronic ground state ($S_0^{min}$) is shown in **Figure 1**, which corresponds to a planar geometry with the $C_s$ symmetry. The current PE dendrimer has 138 harmonic vibrational modes, all vibrational frequencies are list in **Table S1** of the Supporting Information. Here, 93 harmonic vibrational modes belong to the symmetry A', which mainly involves the in-plane molecular vibrations, including the C≡C stretching, C-H stretching, and skeletal bending in molecular plane, etc. The rest of 45 harmonic vibrational modes are of the A" symmetry, which are mainly characterized by the out-of-plane molecular vibrations, such as the skeletal bending, etc. The vibrational modes of symmetry A' display the frequencies within the range of 0.0016-0.4026 eV.

While the A" vibrational modes are mostly low-frequency, and the highest vibrational frequency is 0.1265 eV.

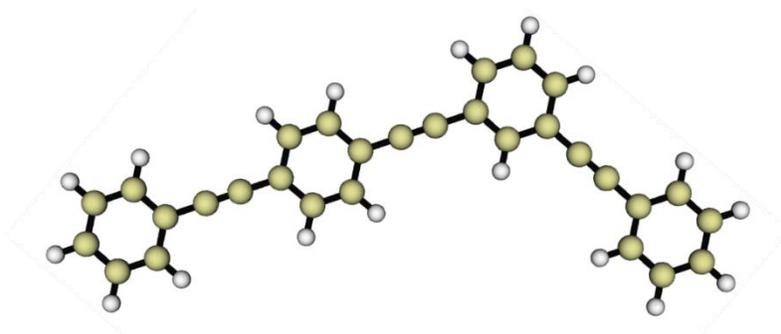

Figure 1. The ground state equilibrium structure of the PE dendrimer. It is composed of two linear PE units (2 rings and 3 rings) linked by meta-substitution. The chartreuse atoms are carbon atoms, and the white atoms represent hydrogen atoms.

The excited-state properties at $S_0^{min}$ are summarized in **Table 1**. In the current PE dendrimer, both the $S_0$-$S_1$ (gs → ππ*) and $S_0$-$S_2$ (gs → ππ*) optical transitions are dipole allowed. The vertical excitation energy (VEE) of the first excited adiabatic state $S_1$ is 3.9784 eV, which is similar to the previous calculations.[36, 37] The adiabatic bright state $S_1$ with large oscillator strength mainly involves the HOMO → LUMO transition, where the involved orbitals are localized on the 3-ring unit as shown in **Figure 2**, this indicates that $S_1$ is a LE state at the 3-ring unit (LE3) within the Franck-Condon region. The VEE of the second adiabatic excited state $S_2$ is 4.5664 eV, which is significantly higher than the corresponding values of $S_1$. This means that the bright state $S_2$ can be excited alone when selecting the appropriate excitation wavelength. The electronic transition of the adiabatic $S_2$ state is mainly characterized by the HOMO-1 → LUMO+1 transition, which corresponds to the local electronic excitation at the 2-ring unit (LE2). While for the third and fourth excited states, $S_3$ and $S_4$ (4.7167 and 4.9360 eV), both are dark states. More information is given in **Figure S1** in the Supporting Information.

Table 1. The VEEs of four low-lying electronic states at the TDDFT/ωB97XD/6-31G* level at $S_0^{min}$ of the PE dendrimer.

| State | VEE (eV) | f | Transition Orbital | Transition |
|---|---|---|---|---|
| $S_1$ | 3.9784 | 2.2540 | HOMO→LUMO (79.2%) | LE3 |
| $S_2$ | 4.5664 | 0.7843 | HOMO-1→LUMO+1 (65.8%) | LE2 |
|  |  |  | HOMO-1→LUMO (12.2%) |  |
| $S_3$ | 4.7167 | 0.0061 | HOMO→LUMO+1 (19.2%) | CT between 2-ring and 3-ring units |
|  |  |  | HOMO-1→LUMO (18.0%) |  |
|  |  |  | HOMO-1→LUMO+2 (13.0%) |  |
| $S_4$ | 4.9360 | 0.0002 | HOMO→LUMO+3 (42.0%) | Local excitation of the middle ring of 3-ring units |
|  |  |  | HOMO-5→LUMO (40.8%) |  |

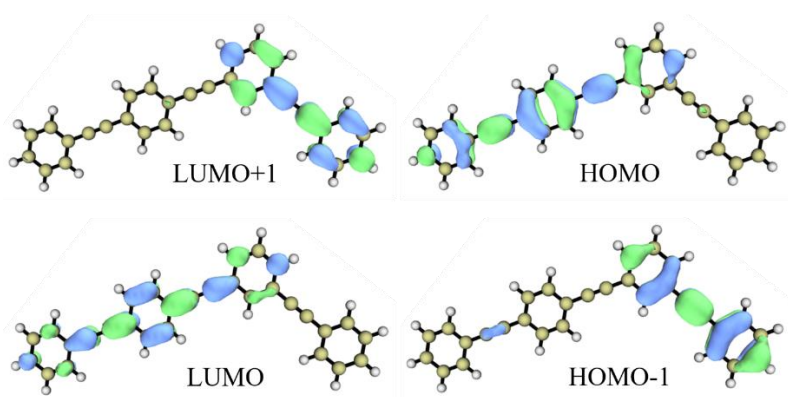

Figure 2. Frontier molecule orbitals at $S_0^{min}$ of the PE dendrimer.

## 3.2. Diabatic Hamiltonian Model

According to the symmetry selection rules, only the A' modes may give the non-zero interstate linear vibronic coupling strengths. We firstly calculated the adiabatic PESs of the $S_1$ and $S_2$ states along each symmetric normal coordinate, the corresponding PESs are shown in **Figure S2**. Among all modes, four dimensionless coordinates $Q_{82}$, $Q_{114}$, $Q_{118}$ and $Q_{120}$ should play potentially important roles in the EET dynamics of the dendrimer, because two potential energy curves along these modes access to each other, and the $S_1/S_2$ avoided crossing region provides the nonadiabatic pathways for the EET dynamics.

The vibrations of modes $v_{82}$, $v_{114}$, $v_{118}$ and $v_{120}$ are shown in **Figure 3**. Here, the modes $v_{82}$ ($\omega_{82}$=0.1439 eV) and $v_{114}$ ($\omega_{114}$=0.2092 eV) mainly involve the vibration of the benzene at the junction, which is the center unit connecting different branches, thus both of them can effectively regulate the interaction between 2-ring and 3-ring units. **Figure 3(a)** shows the in-plane skeletal stretching of mode $v_{82}$, and **Figure 3(b)** describes the in-plane skeletal bending of mode $v_{114}$.

Different to the modes $v_{82}$ and $v_{114}$, the modes $v_{118}$ and $v_{120}$ mainly characterize the local vibration in different branches, which can directly modify the energies of the LE states located at the 2-ring and 3-ring units. For the mode $v_{118}$, the vibrational frequency is 0.2935 eV, and the vibrational mode mainly involves the stretching motions of two C≡C bonds in the 3-ring unit as shown in **Figure 3(c)**. For the mode $v_{120}$, the vibronic frequency is 0.2945 eV, and the vibronic characteristic is the stretching motion of the C≡C bond in the 2-ring unit as shown in **Figure 3(d)**. We also noticed that modes $v_{118}$ and $v_{120}$ should give the largest vibronic coupling terms for the diabatic LE3 and LE2 states, respectively.

Overall, all these four local vibronic modes showing the visible vibronic couplings may couple to the local excited electronic states, which can mediate the ultrafast EET process in the PE dendrimer. Particularly, it is clear that the mode $v_{120}$ is the most important one, and the result here is also highly consistent with some previous works,[36, 37, 47, 52] in which the C≡C bond stretching motion in the 2-ring unit is found to drive the nonadiabatic transition from the adiabatic $S_2$ to $S_1$ states. Thus, these four modes are regarded as the "dominant" ones in the current work, and their effects will be the focus of the following research.

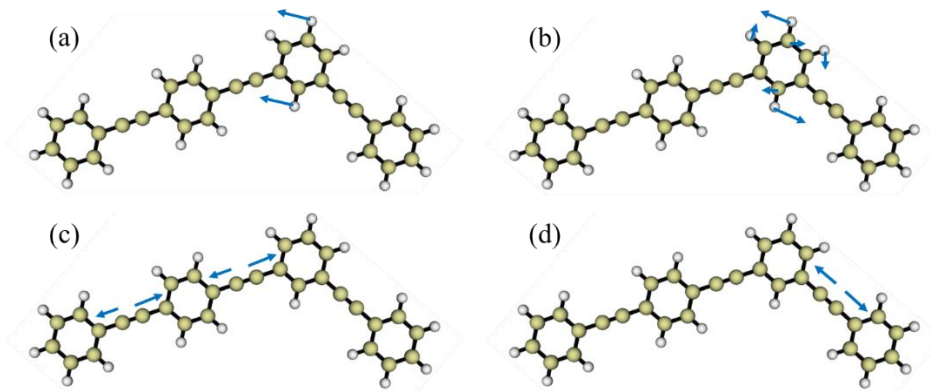

Figure 3. The vibration of four modes, modes (a)$v_{82}$, (b)$v_{114}$, (c)$v_{118}$ and (d)$v_{120}$.

It needs to be emphasized that the second-order vibronic coupling terms are non-neglectable in this work. First, all modes should contribute the second-order coupling terms according to the symmetry selection rule. Second, if we only included the linear vibronic coupling terms (**Figure S2**), it is not easy to obtain the reasonable adiabatic energies based on the diabatic Hamiltonian with respect to the electronic structure calculations results. Only by adding the second-order coupling terms in the diabatic model, the adiabatic PES can be obtained properly. Thus, the final diabatic model is built by following the construction strategy described in the Section 2.2. All fitted parameters of the diabatic model Hamiltonian are collected in **Table S2** of the Supporting Information. In the present model, the RMSEs of four dominant modes (modes $v_{82}$, $v_{114}$, $v_{118}$ and $v_{120}$) are only less than 0.08 eV, and the other RMSEs are even smaller and less than 0.005 eV. Furthermore, the least-squares fit yields the green $S_1$ and yellow $S_2$ curves in **Figure 4** and **Figure S3** of the Supporting Information, which are consistent with the original ab initio PESs as shown in the blue dash-dot and dotted lines therein. All these results demonstrate the current fitting parameters have high precision, and the model itself is reasonable.

The parameter values in the Hamiltonian model further confirm the importance of the above four modes as shown in **Table 2**. The vibronic coupling values of the modes $v_{82}$, $v_{114}$, $v_{118}$ and $v_{120}$ are very large. The $\kappa_i^{(1)}$ value of the mode $v_{118}$ is much larger than the corresponding $\kappa_i^{(2)}$, again indicating this mode gives the strong vibronic coupling for the LE3 state. The opposite is true for the mode $v_{120}$, showing its dominant role in the diabatic LE2 state.

Table 2. The parameter values of the Hamiltonian model of four main modes.

| Parameter | Value/eV | Parameter | Value/eV | Parameter | Value/eV | | | |
|---|---|---|---|---|---|---|---|---|
| $V_{11}^{(0)}$ | 3.9852 | $V_{22}^{(0)}$ | 4.5586 | $\lambda$ | 0.0656 | | | |

| Mode $i$ | Symm | $\omega_i$/eV | $\kappa_i^{(1)}$/eV | $\kappa_i^{(2)}$/eV | $\gamma_i^{(1)}$/eV | $\gamma_i^{(2)}$/eV | $\kappa_i^{(1)}/\omega_i$ | $\kappa_i^{(2)}/\omega_i$ | $\gamma_i^{(1)}/\omega_i$ | $\gamma_i^{(2)}/\omega_i$ |
|---|---|---|---|---|---|---|---|---|---|---|
| 82 | A' | 0.1439 | 0.0709 | -0.0536 | 0.0020 | -0.0062 | 0.4927 | -0.3725 | 0.0139 | -0.0431 |
| 114 | A' | 0.2092 | 0.0923 | -0.0860 | -0.0015 | -0.0185 | 0.4412 | -0.4111 | -0.0072 | -0.0884 |
| 118 | A' | 0.2935 | 0.2175 | -0.0614 | 0.0152 | 0.0699 | 0.7411 | -0.2092 | 0.0518 | 0.2382 |
| 120 | A' | 0.2945 | -0.0827 | -0.3230 | 0.0203 | -0.0197 | -0.2808 | -1.0968 | 0.0689 | -0.0669 |

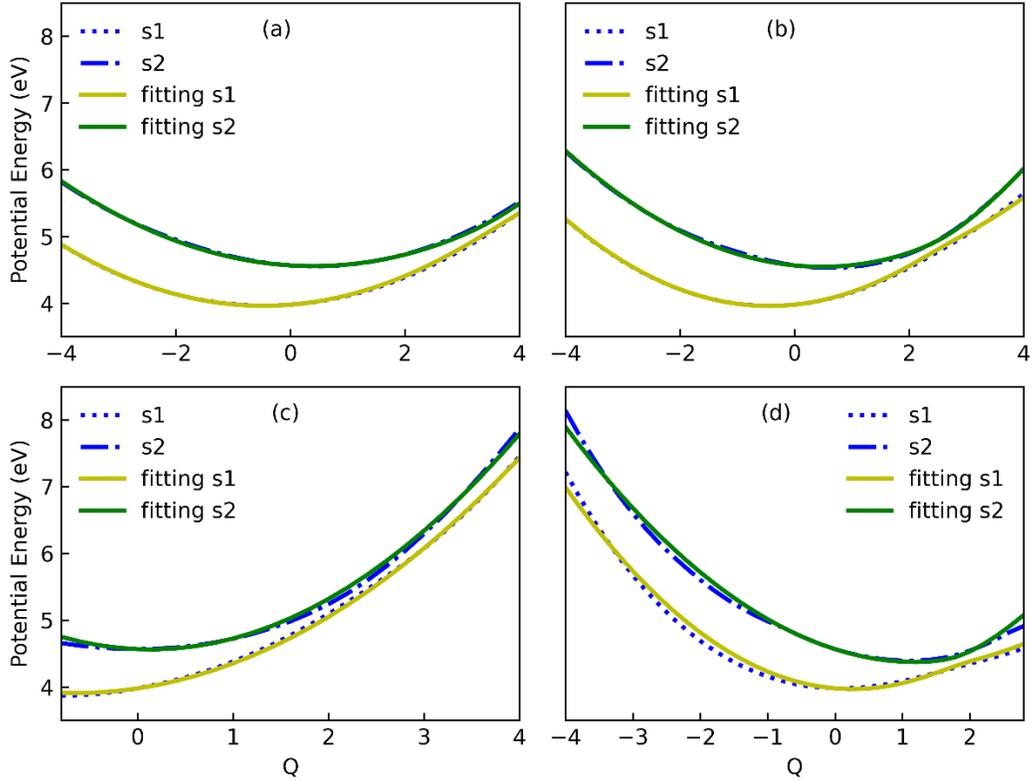

Figure 4. The adiabatic potential energy surfaces of modes (a)$v_{82}$, (b)$v_{114}$, (c)$v_{118}$ and (d)$v_{120}$ compared with the PES from ab initio calculation. The blue dash-dot lines represent the PESs of the S$_2$ state obtained by ab initio calculation, the blue dotted lines represent the PESs of the S$_1$ state obtained by ab initio calculation, the green and yellow lines represent the PESs of both S$_2$ and S$_1$ state obtained by the diagonalization of the diabatic model, respectively. Different horizontal axes were employed for each subfigure to give the better view on the crossing regions.

In this work, the low-frequency modes with frequencies less than 0.01 eV are not considered in the diabatic Hamiltonian models. The main reason considered is that these low-frequency modes cannot be accurately described by the normal mode approximation. In addition, they display the rather different time scales with respect to the ultrafast EET rate. This is also confirmed by the previous work,[50] which has shown that the EET process is mainly related to the high-frequency instantaneous normal modes representing the C≡C stretching motion and the C-H stretching motion of benzene rings. Thus, we do not include the low-frequency modes in the diabatic Hamiltonian. Finally, the largest model includes 129 modes (90 A' and 39 A"). In the future, the contribution of the low-frequency modes can be examined by the consideration of the static disorder.

### 3.3. Dynamics

As the previous analysis based on the PES, modes $v_{82}$, $v_{114}$, $v_{118}$ and $v_{120}$ are inferred to the "dominant" active modes in the current PE dendrimer. Thus, these four modes were firstly included in the reduced model to explore the main characteristics of the EET process. Furthermore, the model Hamiltonian constructed here was truncated into different forms to investigate the linear and second-order vibronic coupling effects.

First of all, it should be important to examine the EET dynamics of the 4-mode diabatic model. **Figure 5(a)** shows the corresponding dynamics results of the optically bright diabatic LE2 state. It is seen that, when only the first-order intrastate coupling is considered in the diabatic Hamiltonian, the early ~10 fs dynamics is entirely determined by the electronic motion. Then, the diabatic LE2 population decays to ~0.15 within 63 fs, which indicates the ultrafast EET dynamics caused by the S$_2$-S$_1$ avoid crossing exists in the current PE dendrimer. This timescale is similar to some previous experimental observations and theoretical simulations.[37, 44, 51]

The frequencies of modes $v_{82}$, $v_{114}$, $v_{118}$ and $v_{120}$ are 0.1439, 0.2092, 0.2935, 0.2945 eV, the corresponding periods are 28.7, 19.8, 14.1, and 14.0 fs, respectively, and the electronic Rabi oscillation is 7.0 fs. Some coherent oscillations are found in the diabatic LE2 population, and due to the limitation of timescale, the further frequency-domain analyses cannot be constructed by Fourier transform. Thus, we only performed roughly counting by eye observations, and the small population oscillation time periods involved are approximately 6, 13, 13, 20, 27 fs, consistent with the electronic Rabi oscillation and the vibrations of modes $v_{82}$, $v_{114}$, $v_{118}$ and $v_{120}$, respectively. Thus, these modes play effective roles on the specific timescale, respectively.

Interestingly, when the second-order intrastate coupling is considered in the model, some distinct dynamics behaviors appear in the time-dependent electronic population (**Figure 5(b)**). Although the early dynamics here is almost identical to the result with only the first-order terms included and the diabatic LE2 population decays rapidly to ~0.37 within 32 fs, here the diabatic LE2 population does not decay further as in the previous case but oscillates around 0.4 until ~160 fs. This manifests the second-order intrastate coupling terms effectively slow down the EET process at some specific timescales. Finally, the diabatic LE2 populations in both cases can reach to ~0.18 at 200 fs. In the whole dynamics, the population recurrence becomes not obvious when the second order coupling terms are included. This phenomenon confirms that the population recurrence patterns are indeed related to the mode vibrations. It is easy to understand such features, because the introduction of the second-order terms can induce the frequency shifts on the excited state for the involved modes.

On this basis, all A' modes with the frequency larger than 0.01 eV were considered in the dynamics to explore their overall effect. The reduced model includes 90 A' modes, which can be roughly divided into four categories based on frequency. Specifically, the first one is the modes with frequencies less than 0.12 eV, whose vibrations mostly involve the in-plane bending vibration of the whole molecule and C≡C bond. The second category is the modes with frequencies between 0.12 and 0.25 eV. Here, most movements are the in-plane bending vibration of the single benzene ring in the molecule, a small part is the in-plane bending vibration of multiple rings or the whole molecule, while the C≡C vibration is not observed in these modes. The third type is the modes with the stretching vibration of the C≡C bond, and their frequencies are range from 0.25 to 0.29 eV. The fourth type is the modes with frequencies large than 0.29 eV, and the vibrations includes the stretching vibration of the C-H bond on the single benzene ring.

**Figure 5** shows the corresponding time-dependent population. As seen, when only the first-order terms were considered, the diabatic LE2 population decays rapidly to ~0.20 within 50 fs, which is consistent to the dynamics of the 4-mode model. Then, the diabatic LE2 population slowly decays to ~0.09 at 130 fs, and finally oscillates around 0.125 until 200 fs (**Figure 5(a)**). Different from the result with the 4-mode model, the diabatic LE2 population recurrence becomes weak in the current situation, which means other 86 modes can damp the quantum coherence of the electronic motions. Similar to the 4-mode model results, when the second-order terms were included in the model, the diabatic LE2 state exhibits the slower population decay, while the population recurrence becomes even weaker, shown in **Figure 5(b)**.

Then, we simulated the dynamics with 129 modes ($\omega_i > 0.01$ eV), which is composed of 90 A' modes and 39 A" modes. As shown in **Figure 5**, the dynamics here is consistent with the results obtained with the 90-mode (A' modes) model. This phenomenon indicates that the A" modes have little influences on the EET process. This is consistent with the basic physical picture. All A" modes should be characterized by the out-of-plane molecular movements, which have the rather low frequencies. Therefore, their time scale is quite different to that for the current ultrafast EET dynamics. In addition, these modes also show the rather small second-order coupling terms, resulting in their weak roles in the diabatic model.

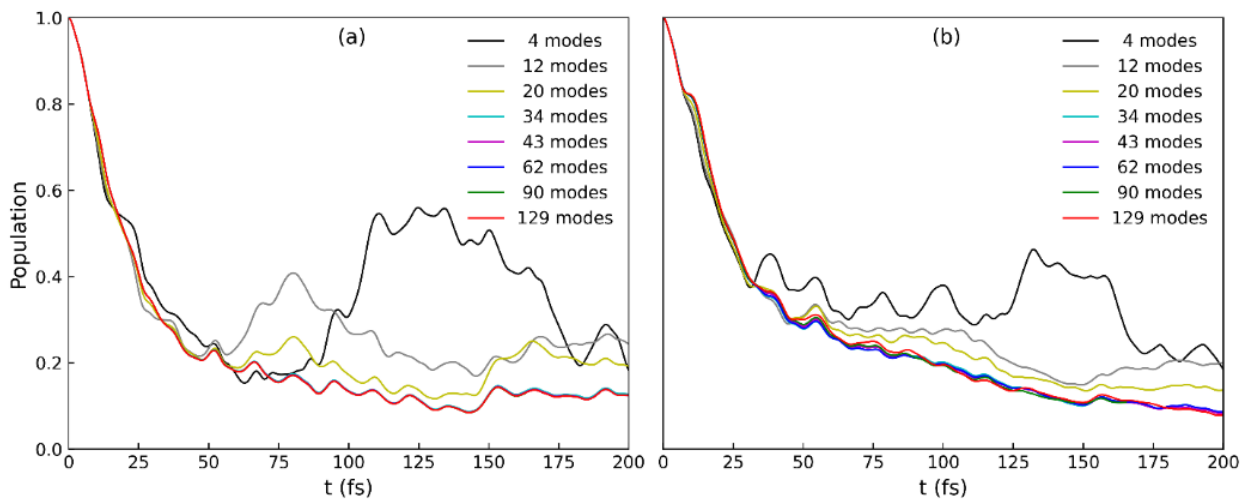

Figure 5. Time-dependent electronic populations of the diabatic LE2 state, which is the initial state, using the model system with 4 modes, 12 modes, 20 modes, 34 modes, 43 modes, 62 modes, 90 modes, 129 modes, respectively. (a) the model containing only the first-order terms, (b) the model containing both the first-order and second-order vibronic coupling terms.

To further explore the influence of other 86 modes on the EET process, some reduced models were constructed according to the electron-vibration coupling strength, *i.e.* $m^{(n)} = \left| \frac{\kappa_i^{(n)}}{\omega_i + 2\gamma_i^{(n)}} \right|$ (n = 1, 2) (as shown in **Figure 6**). The truncation values selected here are 0.45, 0.2, 0.1, 0.05 and 0.01, thus, the reduced models include 12, 20, 34, 43, 62 modes with A' symmetry, respectively, and the corresponding dynamics is shown in **Figure 5**.

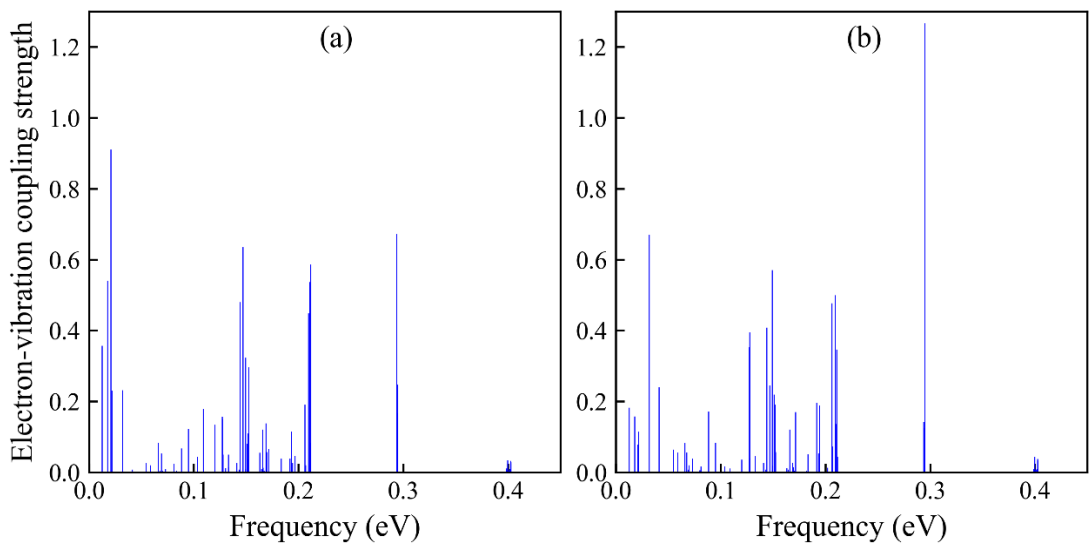

Figure 6. The electron-vibration coupling strength $m^{(n)}$ for different diabatic states. (a) LE1, (b) LE2.

As seen, when only the first-order terms were considered (**Figure 5(a)**), all models show a very similar dynamics in the early time domain, which indicates that the early-time dynamics is fully dominant by the electronic motion and four important modes, while other 86 modes play little roles. After ~50 fs, the obvious population recurrence appears. Here the large population recurrences in the dynamics of the 4-mode model are due to the Poincaré recurrence,[59, 78, 128-134] which has been noted in many reduced model dynamics. While the recurrence can be quickly removed when the model includes 20 vibrational modes. As the contrast, the small-amplitude population oscillation gradually become slow when the model includes more vibrational modes. Here, the models with 12, 20 and 34 modes exhibit different dynamics behaviors from the 4-mode one. When more than 34 modes were considered in the model, the time-dependent diabatic LE2 populations become convergence, which means the mode with $m^{(n)}$ < 0.05 acts as a bystander in the EET dynamics of the current PE system.

The roles of the second-order terms were explained in **Figure 5(b)**. The stronger damping effects on the population recurrence are observed. Similarly, when more modes are included, the coherence is suppressed. The 34-mode, 90-mode and 129-mode Hamiltonian models give almost identical dynamics. This further confirms that, even if the second-order couplings are considered, both A' modes with low vibronic coupling terms ($m^{(n)} < 0.05$) and A" modes do not play visible roles in the current EET dynamics. Certainly, this conclusion is only valid for the current model that does not include the anharmonic effects and mode-mode coupling terms.

The time-dependent electronic coherence (the off-diagonal elements of the reduced density matrix) are shown in **Figure 7**. It is clear that the inclusion of more modes clearly suppresses the quantum coherence motion. In addition, the introduction of the second-order terms gives the rather significant decoherence effects here. After 34 modes are included, the quantum coherences basically remain unchanged, which is consistent with the observation of the population dynamics.

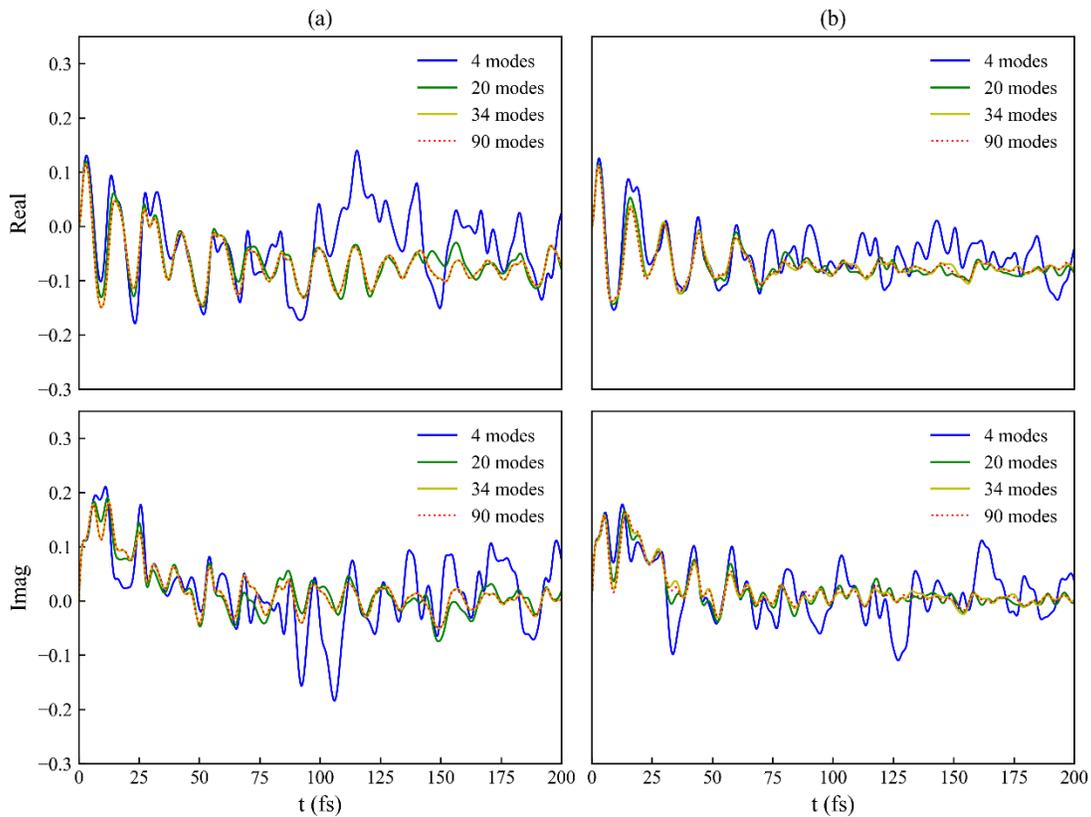

Figure 7. The coherences of the 4-mode, 20-mode, 34-mode and 90-mode models. (a) the model containing only first-order terms, (b) the model containing first-order terms and second-order terms.

## 3.4. Convergence Test

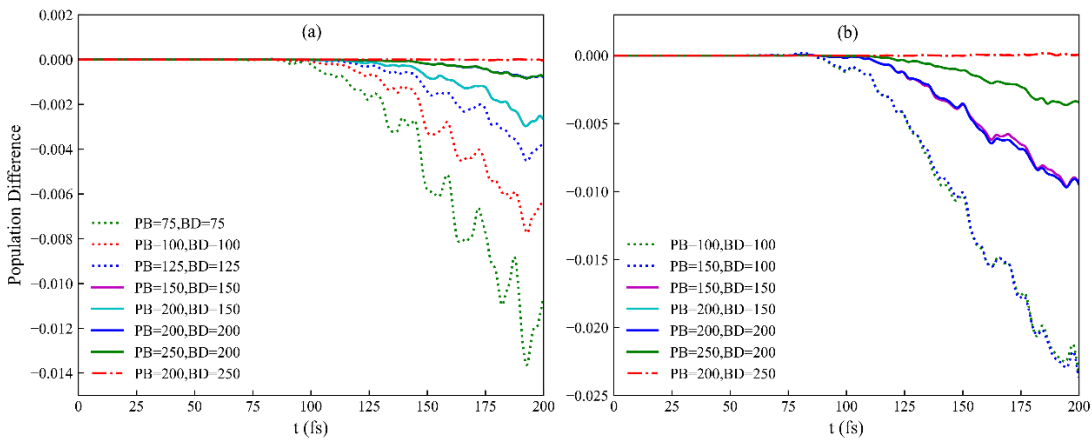

Figure 8. Convergence test in the time-dependent electronic population of the diabatic LE2 state (the initial state) for the 129-mode model, containing only first-order terms (a) and also the second-order terms (b). The results with PB = 250 and BD = 250 are taken as the reference values.

To ensure the accuracy of the dynamics results, we also performed convergence tests, mainly for two parameters PB and BD, which control the size of the MPS, the corresponding results are shown in **Figure 8**. The larger of the matrix product, the higher of the numerical accuracy in the simulation is achieved, while the calculation time is greatly increased. The convergence tests help us to find the appropriate values to achieve the balance between numerical accuracy and calculation time. By comparing the dynamics results of different PB and BD values, the convergence is certainly achieved for the current work. To 129-mode model system containing the first-order terms, the values of PB and BD are 200 and 250, which is adapted in the 129-mode model containing the first-order terms and second-order terms.

## 3.5. Discussion

It is no doubted that there were many excellent dynamics works in the PE dendrimer, which greatly deepens our understandings in the EET process therein. For instance, Roitberg, Fernandez-Alberti, Tretiak, and co-workers[44, 51] once performed the on-the-fly nonadiabatic molecular dynamics simulation with quantum transitions based on the Tully's stochastic fewest-switches algorithm at the semiempirical AM1/CIS level to study the intramolecular EET in the PE dendrimer systems. Their works also clearly demonstrate that the EET process from the adiabatic $S_2$ to $S_1$ states takes place within 50~70 fs. In addition, Lan[37] et al. simulated the on-the-fly surface-hopping nonadiabatic dynamics at TDDFT/CAM-B3LYP/6-31G level and the time-resolved polarization-sensitive transient absorption pump−probe spectra of the PE dendrimer. The result also shows that the adiabatic $S_2$ dynamic is an ultrafast EET process. The advantages of these works are that the EET dynamics was simulated based on the realistic molecular system within the all-atomic models. In this sense, all molecular interactions, such as vibronic couplings, anharmonicity and the couplings between different degrees of freedom, were considered. However, it is clear that the TSH dynamics suffers from some deficiencies, such as the improper treatment of the electronic coherence. Alternatively, it is also possible to build the reduced models with rather limited numbers of vibrational modes and run the quantum dynamics. This provided a better solution of the dynamics part, while some molecular interactions were missing. In this sense, these two types of studies provide the complementary views on the same topic, which is similar to examine two sides of the same coin. This certainly improves our understanding on the EET dynamics of the PE systems. In this sense, our work largely extends the description ability of the second approaches by building the full-dimensional diabatic model to include almost all involved vibrational modes and using the high-efficient tensor network method to simulate the full quantum dynamics. Particularly, we clearly demonstrated that the dominant role of a few modes, such as the C=C bond stretching motions, and the weak roles of other modes. This observation is consistent with these previous studies, while we can also identify other important features, such as different roles of various vibrational modes, the possible influence of the second order coupling terms and the damping effects caused by the dissipations. This certainly improves our understanding of this topic. In addition, this work also demonstrated that the tensor network with the train tensor form gives a very efficient description of the full quantum dynamics for the nonadiabatic process of complicated systems.

## 4. Conclusion

In this article, we have investigated the EET process of the PE dendrimer after excitation to the diabatic LE2 electronic state. The diabatic vibronic model Hamiltonian was considered, which was built based on the electronic structure calculation at the TDDFT/ωB97XD/6-31G* level. This provided us a reliable 2-state diabatic model with 129 modes (90 A' and 39 A").

The quantum chemistry results indicate that four modes, modes $v_{82}$, $v_{114}$, $v_{118}$ and $v_{120}$ with visible linear vibronic coupling terms, may play potential roles on the EET dynamics. Along them, the potential energy curves access to each other, resulting in the $S_1/S_2$ avoided crossing regions and providing the nonadiabatic decay pathways for the EET dynamics.

The EET dynamics starting from the diabatic LE2 was simulated using the tensor network method with the train tensor form. Here, we performed the quantum dynamics simulation of several models, including 4, 90, 129 modes and others. In all cases, the EET in the current PE dendrimer is an ultrafast process. The above four modes with A' symmetry play major roles in the EET process. Particularly, the dominant role of stretching motion of the C≡C bond in the 2-ring part is addressed. Besides, the other 86 modes with A' symmetry can damp the quantum coherence of the electronic motion, while the modes with A" symmetry have little influences on the EET process. Moreover, the inclusion of the second-order intrastate coupling terms can effectively slow down the EET process in the early stage of the dynamics, while they also tend to suppress the quantum coherence.

By combing the electronic structure calculations, the diabatic Hamiltonian construction and the full quantum dynamics simulation with the tensor network methods, this work deepens the understanding of the EET dynamics of PE dendritic macromolecules, which is helpful to improve the energy transfer efficiency of organic photovoltaic cells. In addition, this work also demonstrates that the tensor network method with the train tensor decomposition is effective, which can be generalized to simulate the nonadiabatic dynamics of the large systems at the full quantum level.

## Associated Content

### Supporting Information

The vibrational frequencies and symmetry of 138 harmonic vibrational modes of PE dendrimer, the Frontier molecule orbitals at $S_0^{min}$ of the PE dendrimer, the fitting potential energy surfaces of 129 modes comparing with obtained by *ab initio* calculation, the parameters of diabatic Hamiltonian model.

## Author Information

### Corresponding Author

Zhenggang Lan − SCNU Environmental Research Institute, Guangdong Provincial Key Laboratory of Chemical Pollution and Environmental Safety & MOE Key Laboratory of Environmental Theoretical Chemistry, South China Normal University, Guangzhou 510006, China; School of Environment, South China Normal University, Guangzhou 510006, China; Email: zhenggang.lan@m.scnu.edu.cn, zhenggang.lan@gmail.com

Jiawei Peng − SCNU Environmental Research Institute, Guangdong Provincial Key Laboratory of Chemical Pollution and Environmental Safety & MOE Key Laboratory of Environmental Theoretical Chemistry, South China Normal University, Guangzhou 510006, China; School of Chemistry, South China Normal University, Guangzhou 510006, China; Email: pengjw@aliyun.com

### Conflicts of Interest

There are no conflicts of interest to declare.

## Acknowledgments

This work is supported by NSFC projects (Grants 21933011, 22333003, 22361132528 and 22173107) and the Opening Project of Key Laboratory of Optoelectronic Chemical Materials and Devices of Ministry of Education, Jianghan University (Grant JDGD-202216). The authors thank the Supercomputing Center, Computer Network Information Center, Chinese Academy of Sciences, and the National Supercomputing Center in SunRising-1 for providing computational resources.**References**

1. Mukamel, S., Photochemistry-trees to trap photons. *Nature* **1997,** *388*, 425-427.
2. Swallen, S. F.; Zhu, Z.; Moore, J. S.; Kopelman, R., Correlated excimer formation and molecular rotational dynamics in phenylacetylene dendrimers. *J. Phys. Chem. B* **2000,** *104*, 3988-3995.
3. Tretiak, S.; Mukamel, S., Density matrix analysis and simulation of electronic excitations in conjugated and aggregated molecules. *Chem. Rev.* **2002,** *102*, 3171−3212.
4. D'Ambruoso, G. D.; McGrath, D. V., Energy harvesting in synthetic dendrimer materials. *Adv. Polym. Sci.* **2008,** *214*, 87-147.
5. Bradshaw, D. S.; Andrews, D. L., Mechanisms of light energy harvesting in dendrimers and hyperbranched polymers. *Polymers* **2011,** *3*, 2053-2077.
6. Ceroni, P.; Venturi, M., Photoactive and electroactive dendrimers: future trends and applications. *Aust. J. Chem.* **2011,** *64*, 131-146.
7. Andrews, D. L., Light harvesting in dendrimer materials: designer photophysics and electrodynamics. *J. Mater. Res.* **2012,** *27*, 627-638.
8. Xu, Z.; Kahr, M.; Walker, K. L.; Wilkins, C. L.; Moore, J. S., Phenylacetylene dendrimers by the divergent, convergent, and double-stage convergent methods. *J. Am. Chem. Soc.* **1994,** *116*, 4537-4550.
9. Swallen, S. F.; Kopelman, R.; Moore, J. S.; Devadoss, C., Dendrimer photoantenna supermolecules: energetic funnels, exciton hopping and correlated excimer formation. *J. Mol. Struct.* **1999,** *485*, 585-597.29 quantum dynamics with the multilayer multiconfigurational time-dependent hartree (ML-MCTDH) method. *J. Phys. Chem. C* **2016,** *120*, 1375−1389.

77. Binder, R.; Polkehn, M.; Ma, T.; Burghardt, I., Ultrafast exciton migration in an HJ-aggregate: potential surfaces and quantum dynamics. *Chem. Phys.* **2017,** *482*, 16-26.

78. Jiang, S.; Zheng, J.; Yi, Y.; Xie, Y.; Yuan, F.; Lan, Z., Ultrafast excited-state energy transfer in DTDCTB dimers embedded in a crystal environment: quantum dynamics with the multilayer multiconfigurational time-dependent hartree method. *J. Phys. Chem. C* **2017,** *121*, 27263−27273.

79. Burghardt, I.; Giri, K.; Worth, G. A., Multimode quantum dynamics using Gaussian wavepackets: the Gaussian-based multiconfiguration time-dependent Hartree (G-MCTDH) method applied to the absorption spectrum of pyrazine. *J. Chem. Phys.* **2008,** *129*, 174104.

80. Giese, K.; Petković, M.; Naundorf, H.; Kühn, O., Multidimensional quantum dynamics and infrared spectroscopy of hydrogen bonds. *Phys. Rep.* **2006,** *430*, 211-276.

81. Mondelo-Martell, M.; Brey, D.; Burghardt, I., Quantum dynamical study of inter-chain exciton transport in a regioregular P3HT model system at finite temperature: HJ vs H-aggregate models. *J. Chem. Phys.* **2022,** *157*, 094108.

82. Blasiak, B.; Brey, D.; Koch, W.; Martinazzo, R.; Burghardt, I., Modelling ultrafast dynamics at a conical intersection with regularized diabatic states: An approach based on multiplicative neural networks. *Chem. Phys.* **2022,** *560*, 111542.

83. Thoss, M.; Wang, H.; Miller, W. H., Self-consistent hybrid approach for complex systems: application to the spin-boson model with Debye spectral density. *J. Chem. Phys.* **2001,** *115*, 2991-3005.

84. Wang, H.; Thoss, M., A multilayer multiconfiguration time-dependent hartree simulation of the reaction-coordinate spin-boson model employing an interaction picture. *J. Chem. Phys.* **2017,** *146*, 124112.

85. Tamura, H.; Burghardt, I.; Tsukada, M., Exciton dissociation at thiophene/fullerene
37

Supporting information for

# Ultrafast Excited-State Energy Transfer in Phenylene Ethynylene Dendrimer: Quantum Dynamics with Tensor Network Method

*Sisi Liu[†], Jiawei Peng[†‡*], Peng Bao[§], Qiang Shi[§], Zhenggang Lan[†*]*

[†]SCNU Environmental Research Institute, Guangdong Provincial Key Laboratory of Chemical Pollution and Environmental Safety & MOE Key Laboratory of Environmental Theoretical Chemistry, South China Normal University, Guangzhou 510006, China and School of Environment, South China Normal University, Guangzhou 510006, China

[‡] School of Chemistry, South China Normal University, Guangzhou 510006, China

[§]Beijing National Laboratory for Molecular Sciences, State Key Laboratory for Structural Chemistry of Unstable and Stable Species, CAS Research/Education Center for Excellence in Molecular Sciences, Institute of Chemistry, Chinese Academy of Sciences, Zhongguancun, Beijing 100190, China and University of Chinese Academy of Sciences, Beijing 100049, China

**E-mail:** zhenggang.lan@m.scnu.edu.cn; pengjw@aliyun.com.

# S1: The vibrational frequencies of 138 harmonic vibrational modes of PE dendrimer.

Table S1. The vibrational frequencies of 138 harmonic vibrational modes of PE dendrimer.

| Mode $i$ | Symm | $\omega_i$/eV | Mode $i$ | Symm | $\omega_i$/eV | Mode $i$ | Symm | $\omega_i$/eV |
|---|---|---|---|---|---|---|---|---|
| 1 | A" | 0.0014 | 47 | A" | 0.0885 | 93 | A' | 0.1655 |
| 2 | A' | 0.0016 | 48 | A" | 0.0885 | 94 | A' | 0.1659 |
| 3 | A" | 0.0018 | 49 | A" | 0.0923 | 95 | A' | 0.1660 |
| 4 | A" | 0.0023 | 50 | A' | 0.0948 | 96 | A' | 0.1668 |
| 5 | A" | 0.0028 | 51 | A" | 0.0973 | 97 | A' | 0.1686 |
| 6 | A' | 0.0038 | 52 | A" | 0.0973 | 98 | A' | 0.1692 |
| 7 | A" | 0.0061 | 53 | A" | 0.1021 | 99 | A' | 0.1701 |
| 8 | A' | 0.0067 | 54 | A' | 0.1035 | 100 | A' | 0.1701 |
| 9 | A" | 0.0071 | 55 | A" | 0.1072 | 101 | A' | 0.1713 |
| 10 | A' | 0.0123 | 56 | A" | 0.1076 | 102 | A' | 0.1818 |
| 11 | A" | 0.0128 | 57 | A" | 0.1081 | 103 | A' | 0.1831 |
| 12 | A' | 0.0175 | 58 | A" | 0.1082 | 104 | A' | 0.1867 |
| 13 | A" | 0.0177 | 59 | A' | 0.1088 | 105 | A' | 0.1867 |
| 14 | A' | 0.0205 | 60 | A" | 0.1156 | 106 | A' | 0.1915 |
| 15 | A" | 0.0214 | 61 | A" | 0.1169 | 107 | A' | 0.1931 |
| 16 | A' | 0.0216 | 62 | A" | 0.1176 | 108 | A' | 0.1940 |
| 17 | A' | 0.0317 | 63 | A" | 0.1176 | 109 | A' | 0.1963 |
| 18 | A" | 0.0333 | 64 | A' | 0.1199 | 110 | A' | 0.2018 |
| 19 | A" | 0.0365 | 65 | A" | 0.1228 | 111 | A' | 0.2059 |
| 20 | A' | 0.0410 | 66 | A" | 0.1229 | 112 | A' | 0.2064 |
| 21 | A" | 0.0428 | 67 | A" | 0.1232 | 113 | A' | 0.2065 |
| 22 | A" | 0.0501 | 68 | A" | 0.1232 | 114 | A' | 0.2092 |
| 23 | A" | 0.0506 | 69 | A" | 0.1247 | 115 | A' | 0.2102 |
| 24 | A" | 0.0514 | 70 | A" | 0.1265 | 116 | A' | 0.2106 |
| 25 | A" | 0.0516 | 71 | A" | 0.1265 | 117 | A' | 0.2113 |
| 26 | A" | 0.0516 | 72 | A' | 0.1271 | 118 | A' | 0.2935 |
| 27 | A' | 0.0545 | 73 | A' | 0.1274 | 119 | A' | 0.2943 |
| 28 | A' | 0.0585 | 74 | A' | 0.1275 | 120 | A' | 0.2945 |
| 29 | A" | 0.0590 | 75 | A' | 0.1302 | 121 | A' | 0.3984 |
| 30 | A' | 0.0627 | 76 | A' | 0.1327 | 122 | A' | 0.3984 |
| 31 | A" | 0.0627 | 77 | A' | 0.1327 | 123 | A' | 0.3994 |
| 32 | A' | 0.0657 | 78 | A' | 0.1392 | 124 | A' | 0.3995 |

| Mode $i$ | Symm | $\omega_i$/eV | Mode $i$ | Symm | $\omega_i$/eV | Mode $i$ | Symm | $\omega_i$/eV |
|---|---|---|---|---|---|---|---|---|
| 33 | A' | 0.0674 | 79 | A' | 0.1393 | 125 | A' | 0.3996 |
| 34 | A" | 0.0683 | 80 | A' | 0.1407 | 126 | A' | 0.4003 |
| 35 | A' | 0.0689 | 81 | A' | 0.1429 | 127 | A' | 0.4004 |
| 36 | A" | 0.0690 | 82 | A' | 0.1439 | 128 | A' | 0.4007 |
| 37 | A' | 0.0695 | 83 | A' | 0.1466 | 129 | A' | 0.4007 |
| 38 | A" | 0.0712 | 84 | A' | 0.1491 | 130 | A' | 0.4015 |
| 39 | A' | 0.0728 | 85 | A' | 0.1492 | 131 | A' | 0.4015 |
| 40 | A" | 0.0776 | 86 | A' | 0.1492 | 132 | A' | 0.4017 |
| 41 | A' | 0.0797 | 87 | A' | 0.1506 | 133 | A' | 0.4023 |
| 42 | A' | 0.0797 | 88 | A' | 0.1517 | 134 | A' | 0.4023 |
| 43 | A' | 0.0808 | 89 | A' | 0.1518 | 135 | A' | 0.4023 |
| 44 | A' | 0.0831 | 90 | A' | 0.1522 | 136 | A' | 0.4024 |
| 45 | A" | 0.0880 | 91 | A' | 0.1628 | 137 | A' | 0.4025 |
| 46 | A' | 0.0882 | 92 | A' | 0.1644 | 138 | A' | 0.4026 |

## S2: Frontier molecule orbitals at $S_0^{min}$ of PE dendrimer.

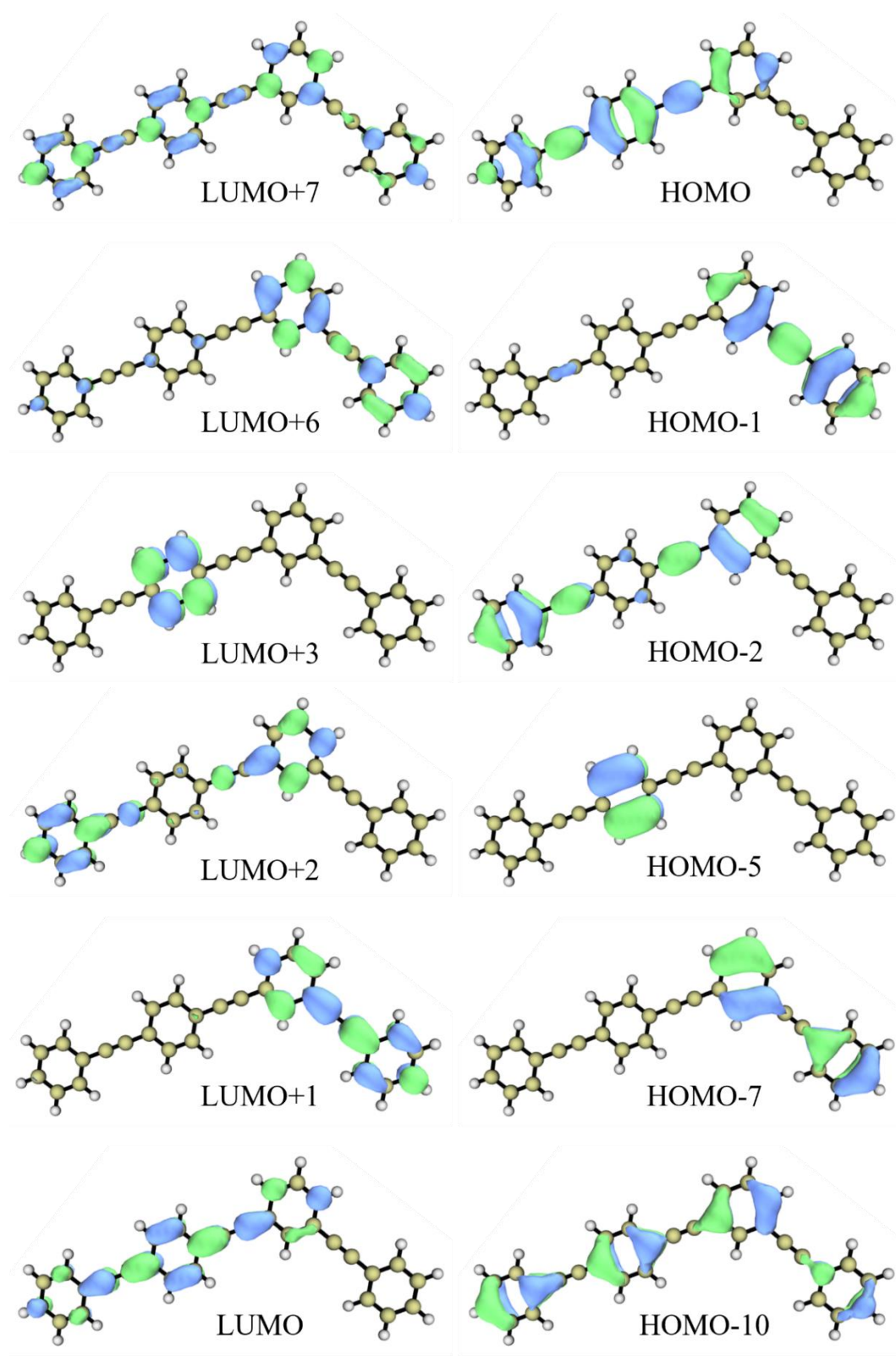

Figure S1. The Frontier molecule orbitals at $S_0^{min}$ of PE dendrimer.

**S3: The potential energy surfaces along the 4 main dimensionless normal coordinates, when only the first-order intrastate couplings are considered.**

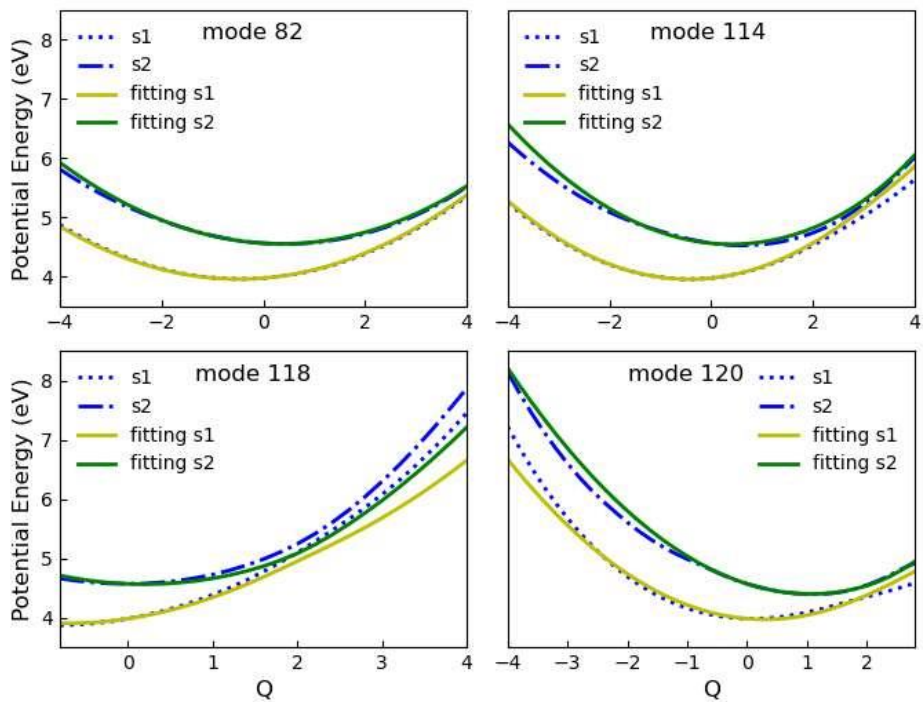

Figure S2. The adiabatic potential energy surfaces of Modes (a)82, (b)114, (c)118 and (d)120 compared with the PES from ab initio calculation, which only the first-order intrastate coupling is considered. The blue dash-dot lines represent the PESs of the S2 state obtained by ab initio calculation, the blue dotted lines represent the PESs of the S1 state obtained by ab initio calculation, the green and yellow lines represent the PESs of both S2 and S1 state obtained by the diagonalization of the diabatic model, respectively. Different horizontal axes were employed for each subfigure to give the better view on the crossing regions.

## S4: The potential energy surfaces of 129 modes after fitting.

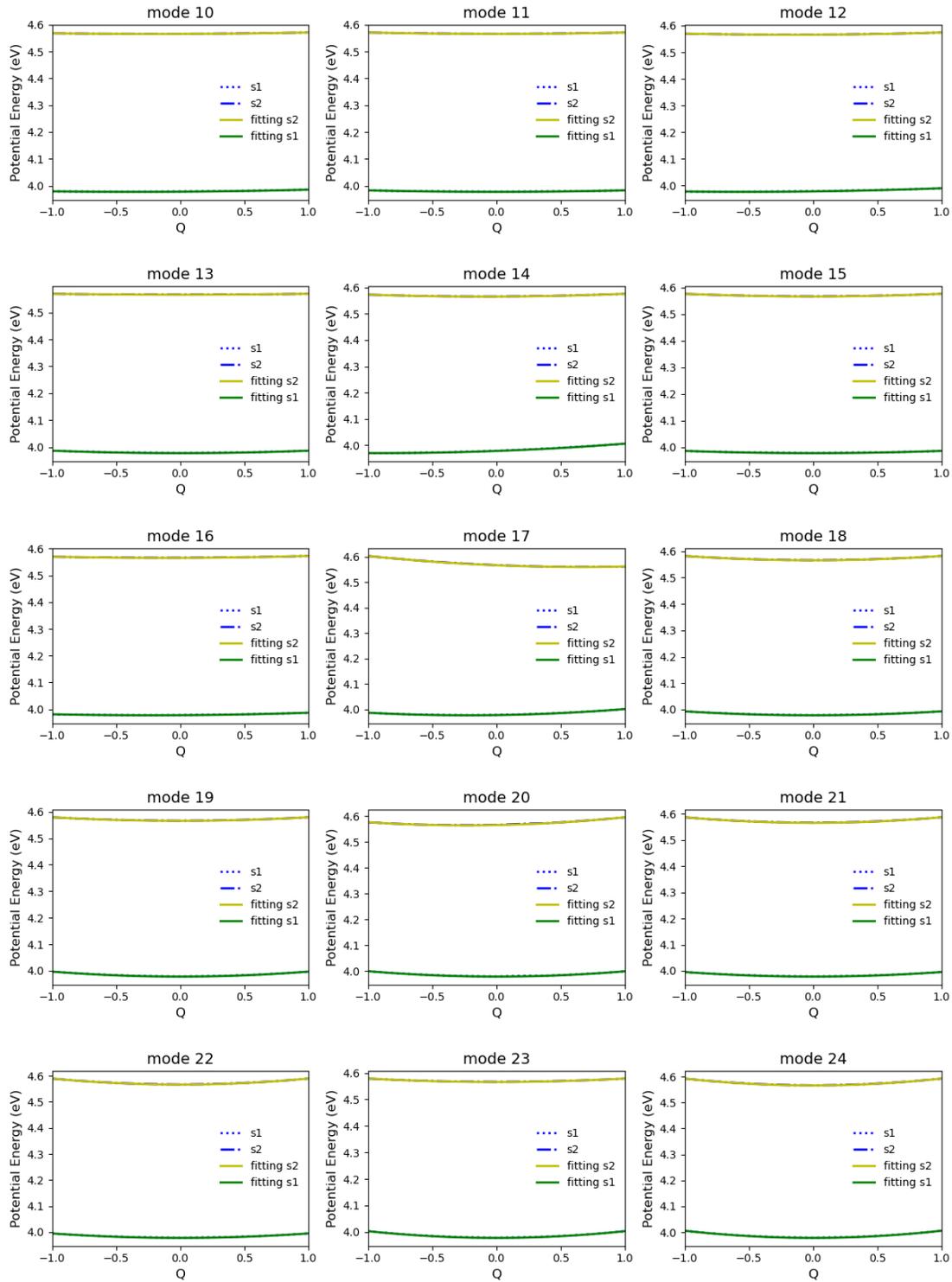

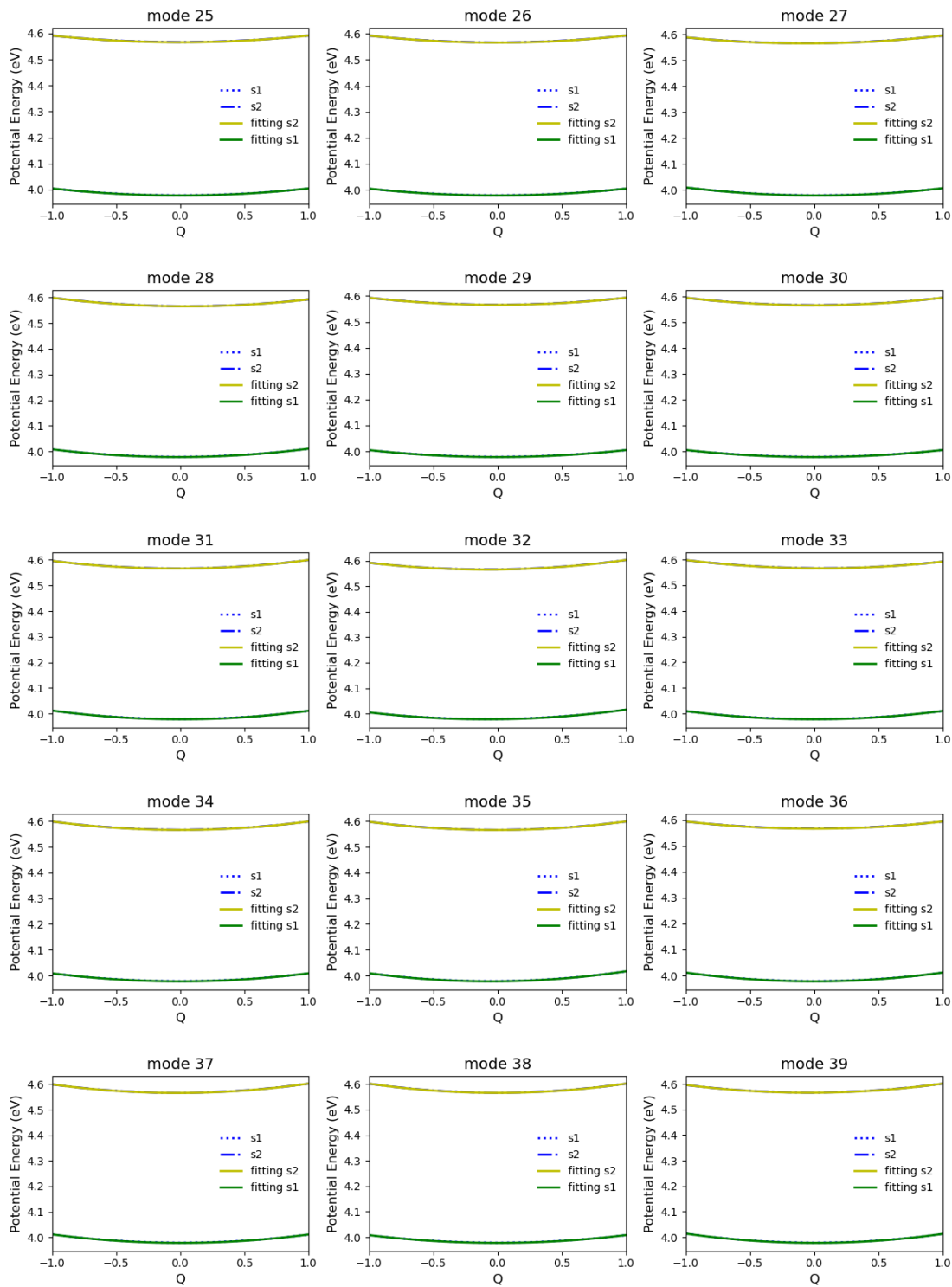

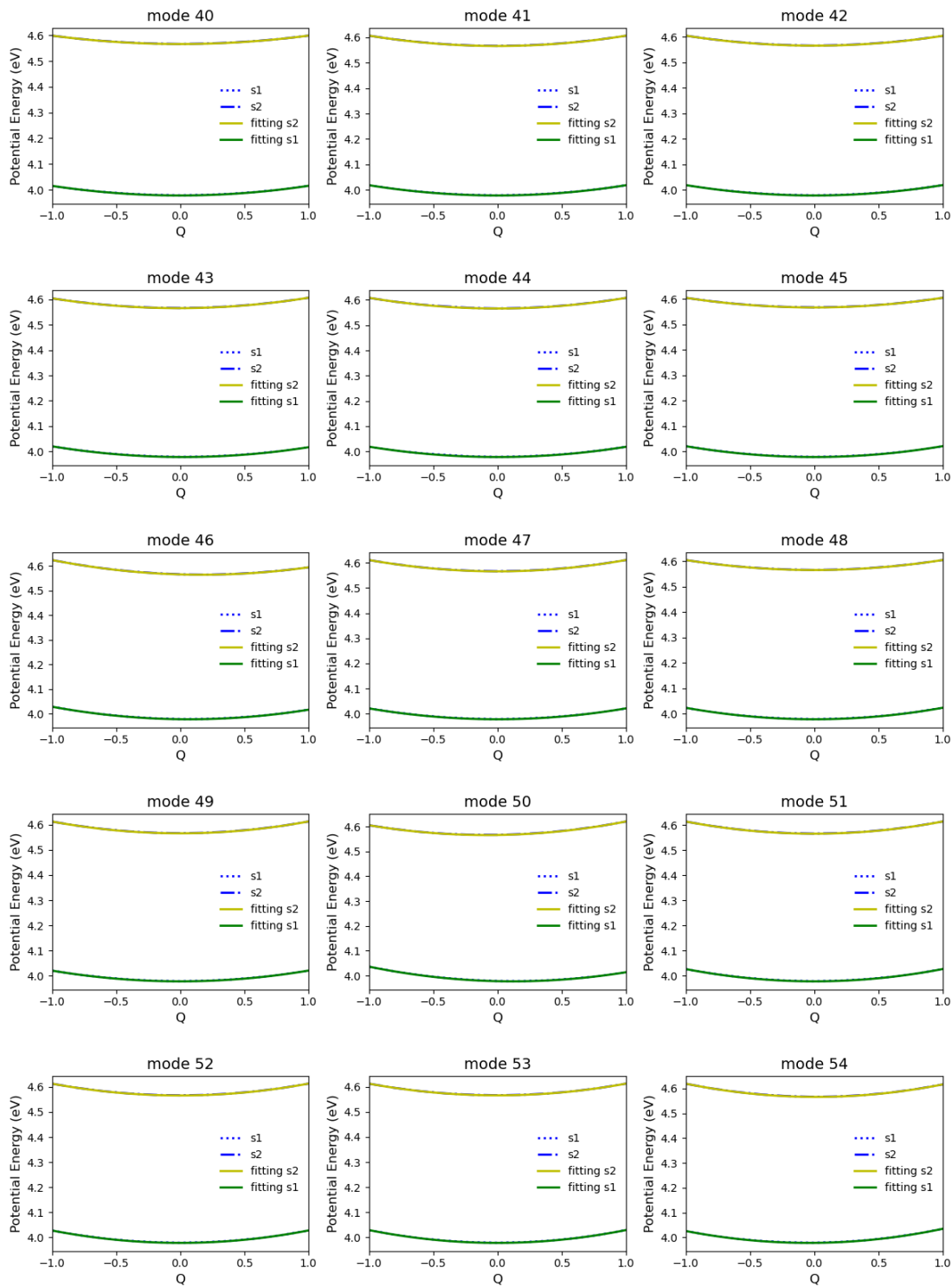

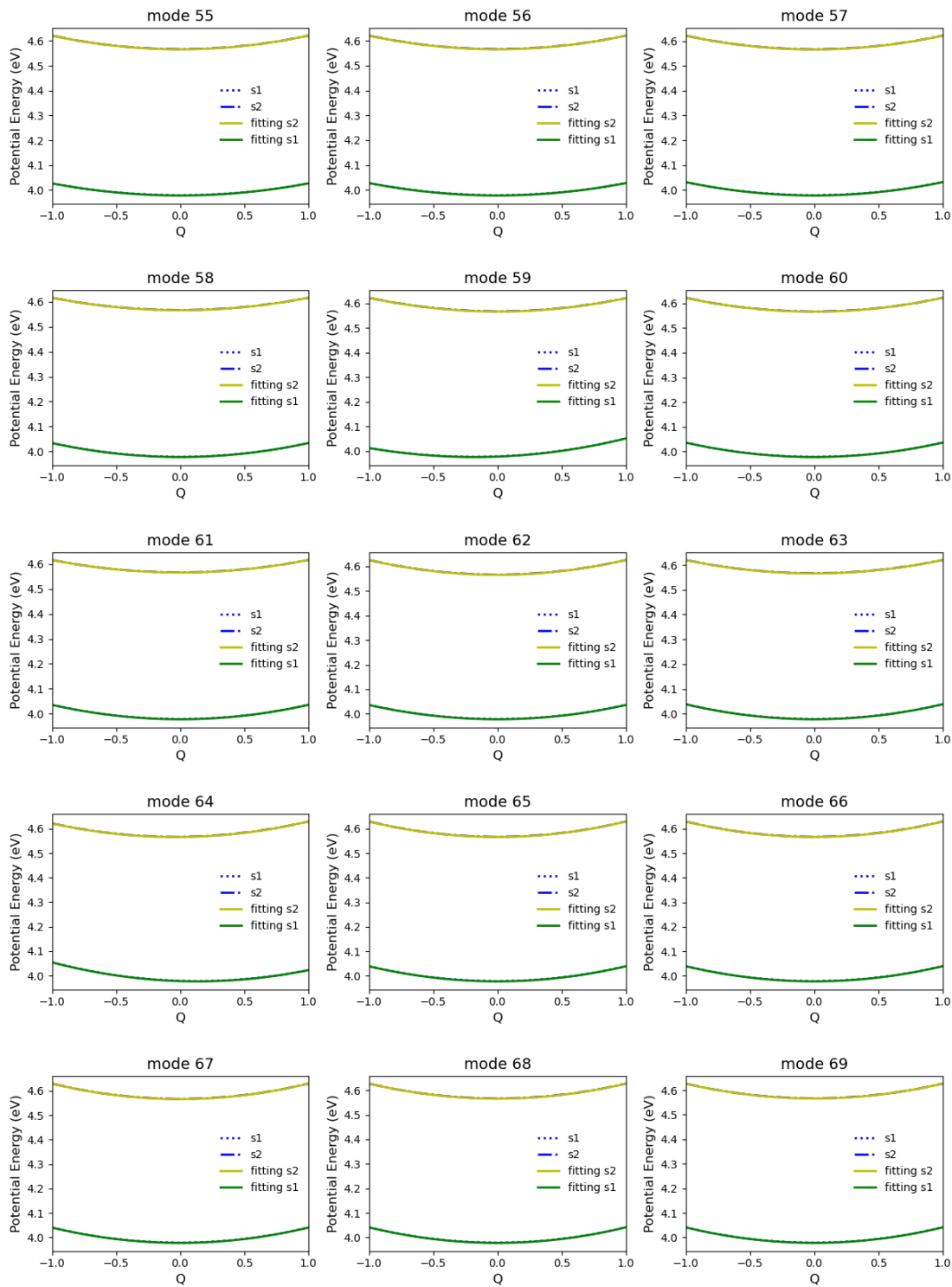

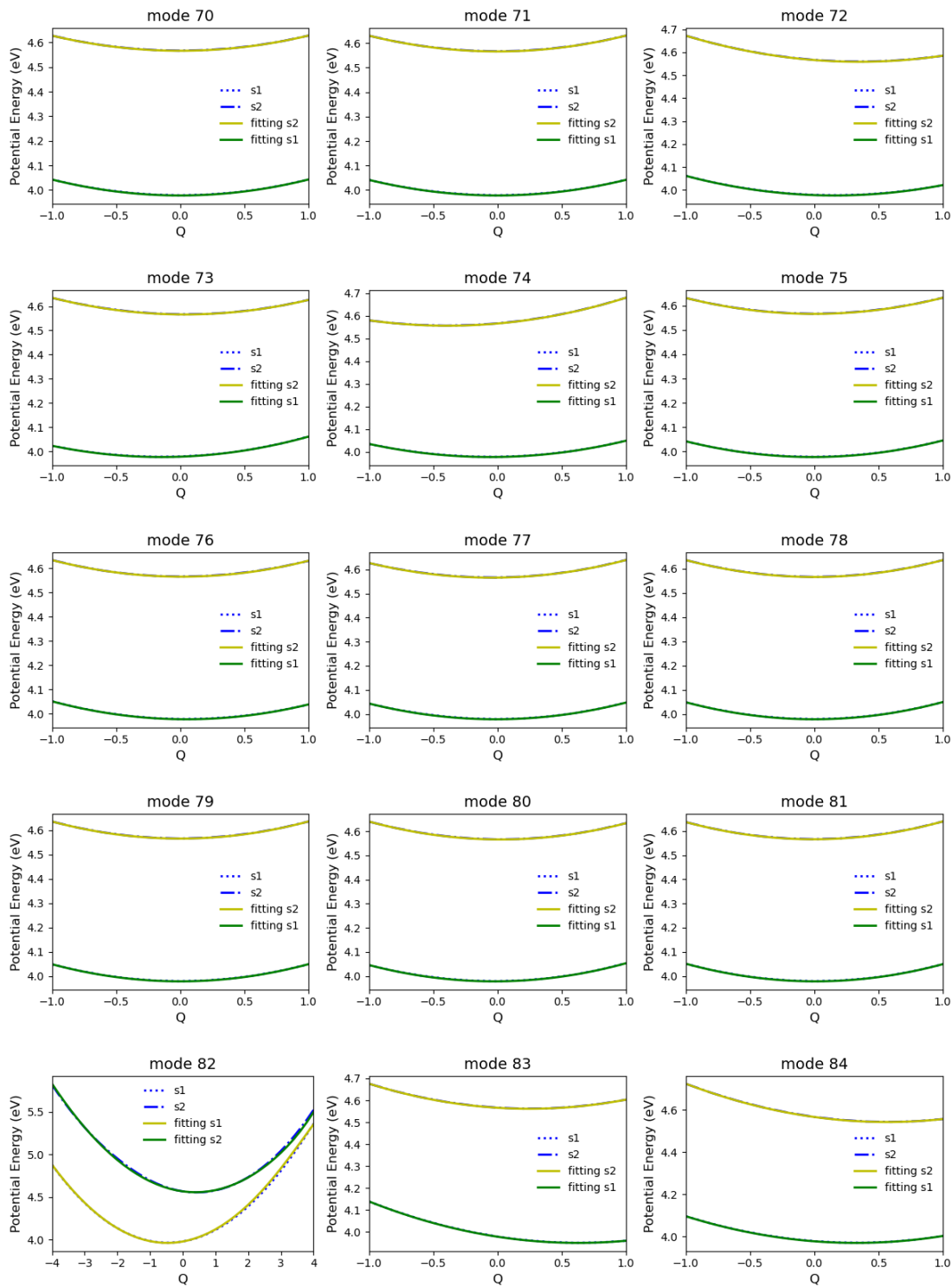
53

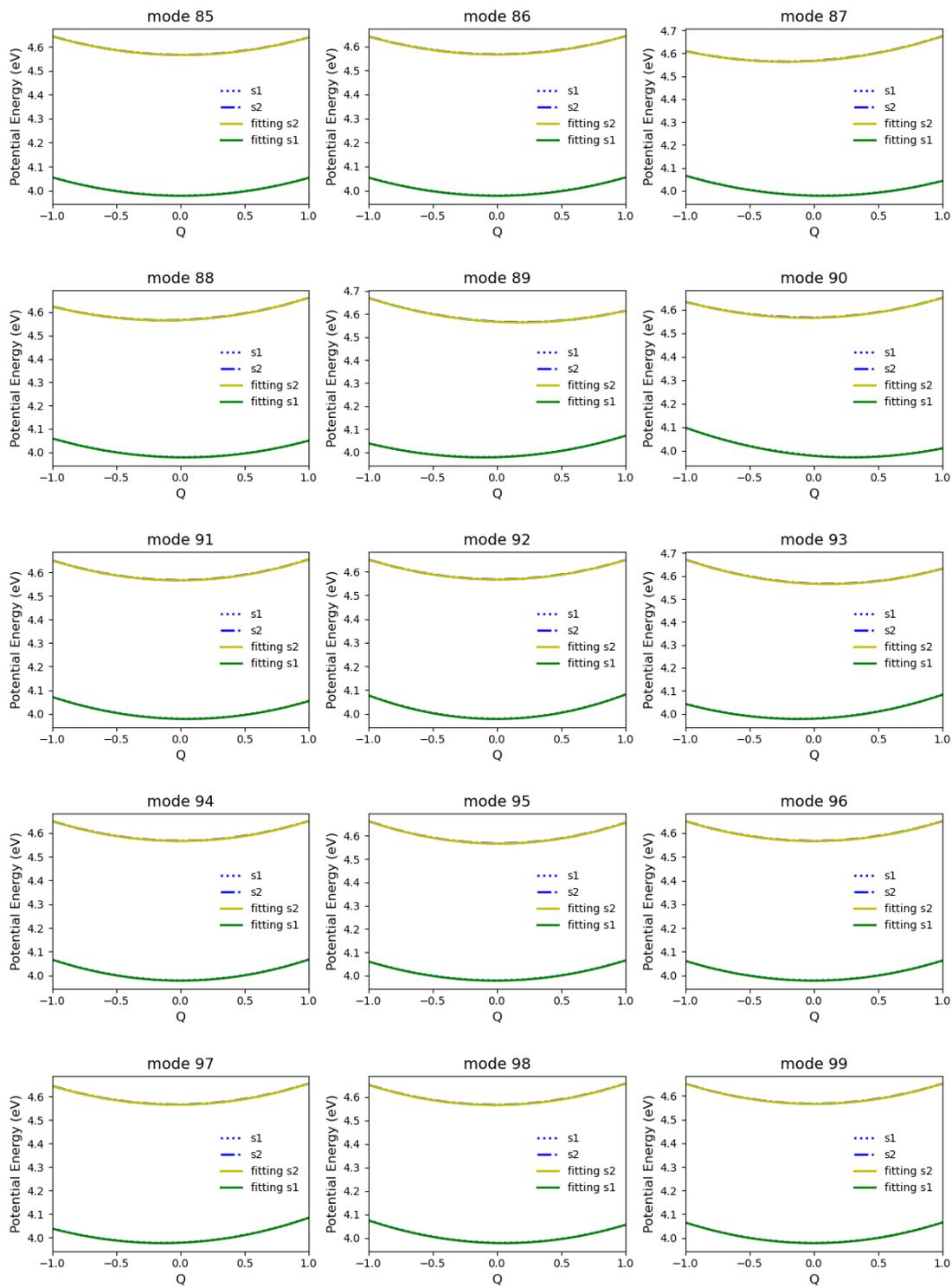

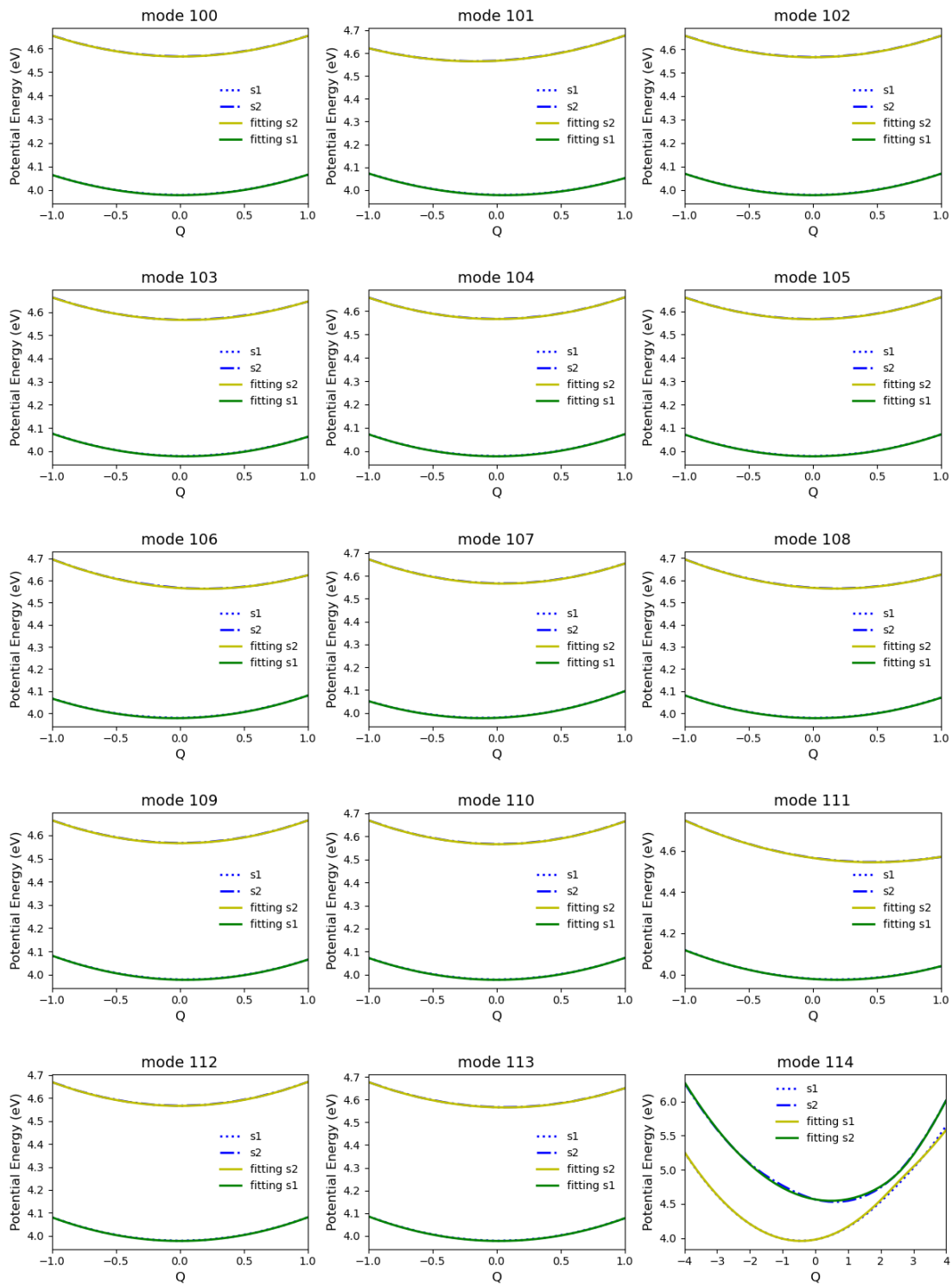

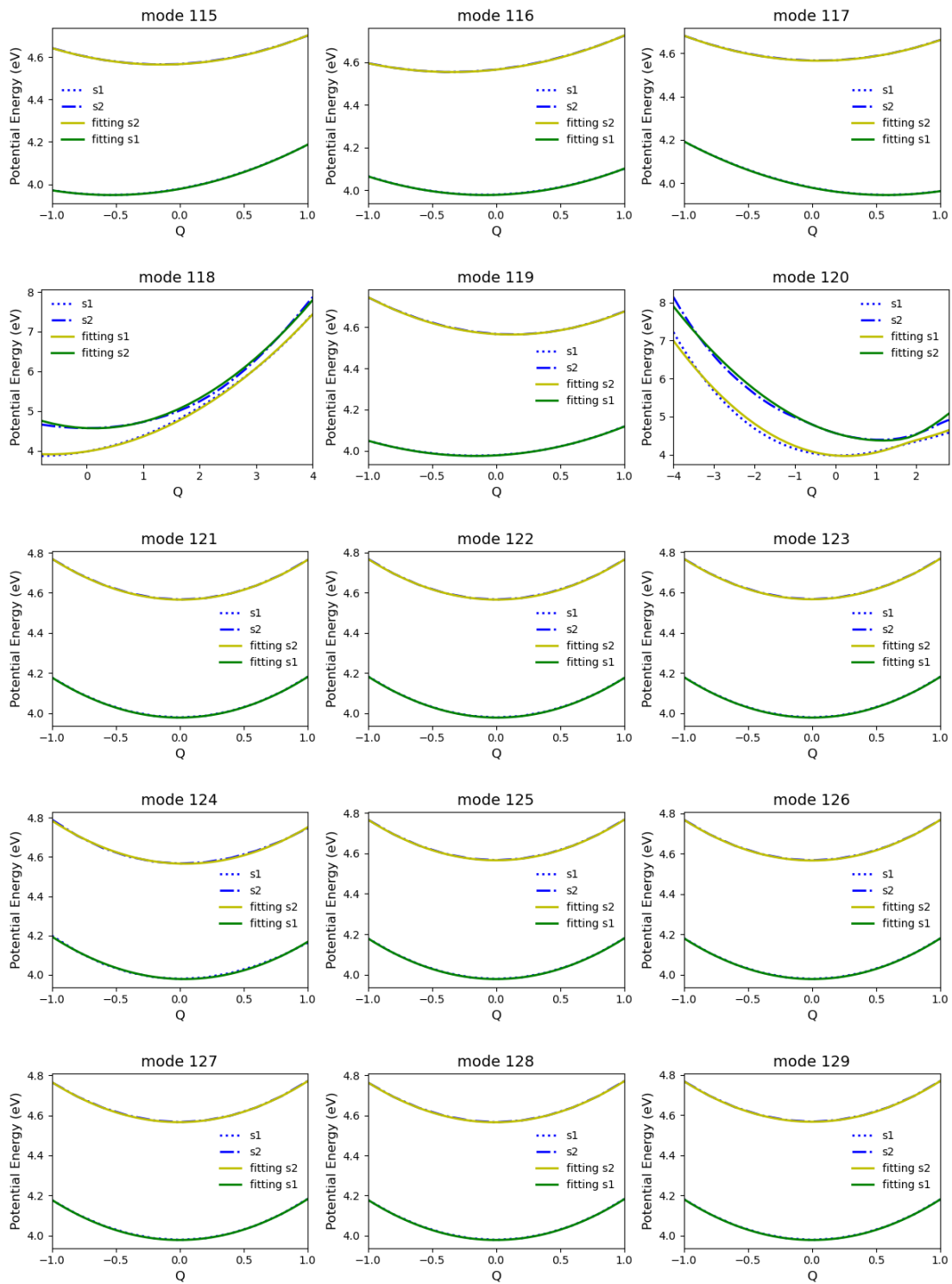

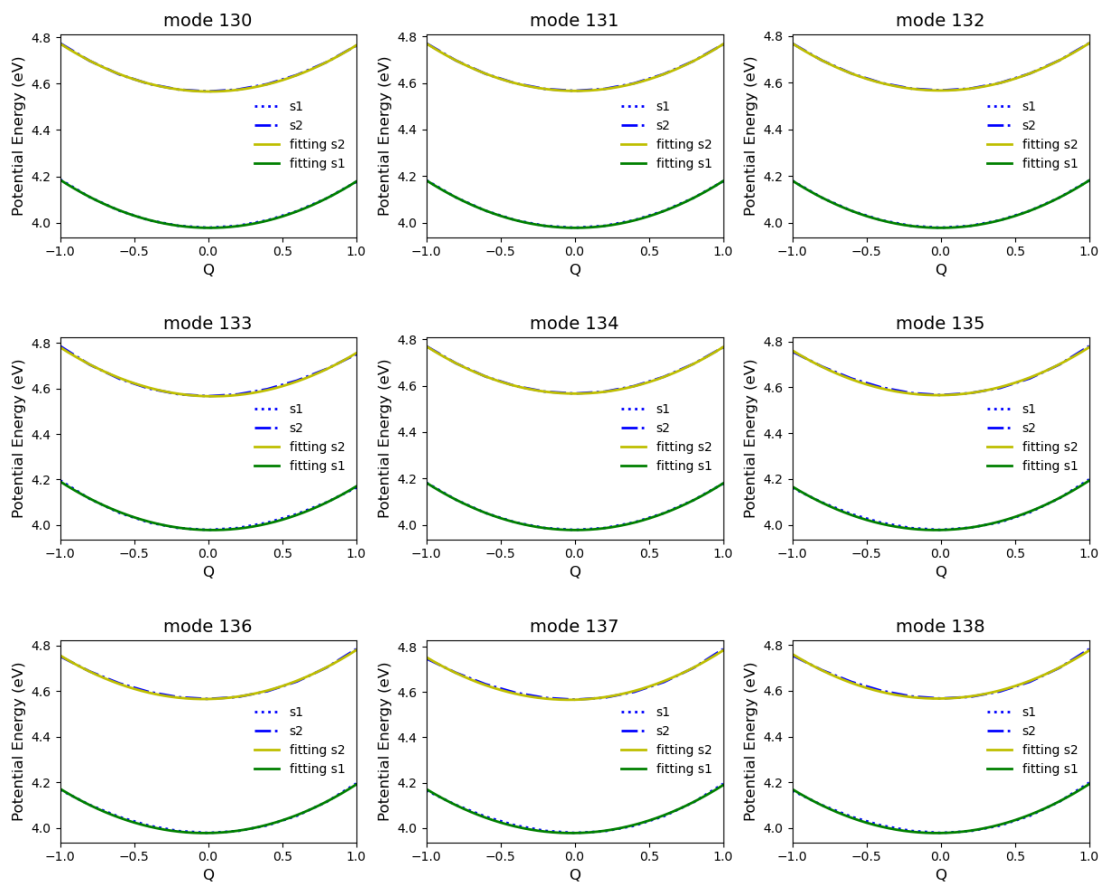

Figure S3. The fitting potential energy surfaces of 129 modes comparing with obtained by *ab initio* calculation, which the first-order intrastate coupling and second-order term are considered. The blue dash-dot lines represent the PESs of the $S_2$ state obtained by ab initio calculation, the blue dotted lines represent the PESs of the $S_1$ state obtained by ab initio calculation, the green and yellow lines represent the PESs of both $S_2$ and $S_1$ state obtained by the diagonalization of the diabatic model, respectively. Different horizontal axes were employed for each subfigure to give the better view on the crossing regions.

## S5: The parameters of diabatic Hamiltonian model

Table S2. The parameters of diabatic Hamiltonian model.

| Mode $i$ | Symm | $\omega_i$/eV | $\kappa_i^{(1)}$/eV | $\kappa_i^{(2)}$/eV | $\gamma_i^{(1)}$/eV | $\gamma_i^{(2)}$/eV | $\kappa_i^{(1)}/\omega_i$ | $\kappa_i^{(2)}/\omega_i$ | $\gamma_i^{(1)}/\omega_i$ | $\gamma_i^{(2)}/\omega_i$ |
|---|---|---|---|---|---|---|---|---|---|---|
| 10 | A' | 0.0123 | 0.0031 | 0.0014 | -0.0018 | -0.0023 | 0.2520 | 0.1138 | -0.1463 | -0.1870 |
| 11 | A" | 0.0128 | 0.0000 | 0.0000 | -0.0010 | -0.0012 | 0.0000 | 0.0000 | -0.0781 | -0.0938 |
| 12 | A' | 0.0175 | 0.0061 | 0.0020 | -0.0031 | -0.0024 | 0.3486 | 0.1143 | -0.1771 | -0.1371 |
| 13 | A" | 0.0177 | 0.0000 | 0.0000 | 0.0003 | -0.0055 | 0.0000 | 0.0000 | 0.0169 | -0.3107 |
| 14 | A' | 0.0205 | 0.0181 | 0.0014 | -0.0003 | -0.0013 | 0.8829 | 0.0683 | -0.0146 | -0.0634 |
| 15 | A" | 0.0214 | 0.0000 | 0.0000 | -0.0024 | -0.0007 | 0.0000 | 0.0000 | -0.1121 | -0.0327 |
| 16 | A' | 0.0216 | 0.0029 | 0.0013 | -0.0045 | -0.0051 | 0.1343 | 0.0602 | -0.2083 | -0.2361 |
| 17 | A' | 0.0317 | 0.0076 | -0.0211 | 0.0006 | -0.0001 | 0.2397 | -0.6656 | 0.0189 | -0.0032 |
| 18 | A" | 0.0333 | 0.0000 | 0.0000 | -0.0017 | -0.0003 | 0.0000 | 0.0000 | -0.0511 | -0.0090 |
| 19 | A" | 0.0365 | 0.0000 | 0.0000 | 0.0007 | -0.0048 | 0.0000 | 0.0000 | 0.0192 | -0.1315 |
| 20 | A' | 0.0410 | -0.0003 | 0.0098 | 0.0002 | 0.0000 | -0.0073 | 0.2390 | 0.0049 | 0.0000 |
| 21 | A" | 0.0428 | 0.0000 | 0.0000 | -0.0039 | 0.0001 | 0.0000 | 0.0000 | -0.0911 | 0.0023 |
| 22 | A" | 0.0501 | 0.0000 | 0.0000 | -0.0081 | -0.0019 | 0.0000 | 0.0000 | -0.1617 | -0.0379 |
| 23 | A" | 0.0506 | 0.0000 | 0.0000 | 0.0003 | -0.0125 | 0.0000 | 0.0000 | 0.0059 | -0.2470 |
| 24 | A" | 0.0514 | 0.0000 | 0.0000 | 0.0020 | 0.0009 | 0.0000 | 0.0000 | 0.0389 | 0.0175 |
| 25 | A" | 0.0516 | 0.0000 | 0.0000 | 0.0013 | -0.0005 | 0.0000 | 0.0000 | 0.0252 | -0.0097 |
| 26 | A" | 0.0516 | 0.0000 | 0.0000 | 0.0006 | 0.0009 | 0.0000 | 0.0000 | 0.0116 | 0.0174 |
| 27 | A' | 0.0545 | -0.0015 | 0.0034 | 0.0024 | -0.0007 | -0.0275 | 0.0624 | 0.0440 | -0.0128 |
| 28 | A' | 0.0585 | 0.0012 | -0.0033 | 0.0020 | 0.0007 | 0.0205 | -0.0564 | 0.0342 | 0.0120 |
| 29 | A" | 0.0590 | 0.0000 | 0.0000 | -0.0027 | -0.0022 | 0.0000 | 0.0000 | -0.0458 | -0.0373 |
| 30 | A' | 0.0627 | 0.0000 | 0.0000 | -0.0042 | -0.0024 | 0.0000 | 0.0000 | -0.0670 | -0.0383 |
| 31 | A" | 0.0627 | 0.0000 | 0.0000 | 0.0021 | 0.0004 | 0.0000 | 0.0000 | 0.0335 | 0.0064 |
| 32 | A' | 0.0657 | 0.0054 | 0.0053 | -0.0006 | -0.0007 | 0.0822 | 0.0807 | -0.0091 | -0.0107 |
| 33 | A' | 0.0674 | 0.0003 | -0.0032 | -0.0015 | -0.0047 | 0.0045 | -0.0475 | -0.0223 | -0.0697 |
| 34 | A" | 0.0683 | 0.0000 | 0.0000 | -0.0030 | -0.0014 | 0.0000 | 0.0000 | -0.0439 | -0.0205 |
| 35 | A' | 0.0689 | 0.0037 | 0.0005 | 0.0008 | -0.0020 | 0.0537 | 0.0073 | 0.0116 | -0.0290 |
| 36 | A" | 0.0690 | 0.0000 | 0.0000 | -0.0006 | -0.0070 | 0.0000 | 0.0000 | -0.0087 | -0.1014 |
| 37 | A' | 0.0695 | -0.0003 | 0.0014 | -0.0017 | 0.0002 | -0.0043 | 0.0201 | -0.0245 | 0.0029 |
| 38 | A" | 0.0712 | 0.0000 | 0.0000 | -0.0052 | 0.0005 | 0.0000 | 0.0000 | -0.0730 | 0.0070 |
| 39 | A' | 0.0728 | -0.0006 | 0.0025 | -0.0005 | -0.0035 | -0.0082 | 0.0343 | -0.0069 | -0.0481 |

| Mode $i$ | Symm | $\omega_i$/eV | $\kappa_i^{(1)}$/eV | $\kappa_i^{(2)}$/eV | $\gamma_i^{(1)}$/eV | $\gamma_i^{(2)}$/eV | $\kappa_i^{(1)}/\omega_i$ | $\kappa_i^{(2)}/\omega_i$ | $\gamma_i^{(1)}/\omega_i$ | $\gamma_i^{(2)}/\omega_i$ |
|---|---|---|---|---|---|---|---|---|---|---|
| 40 | A" | 0.0776 | 0.0000 | 0.0000 | -0.0013 | -0.0062 | 0.0000 | 0.0000 | -0.0168 | -0.0799 |
| 41 | A' | 0.0797 | -0.0001 | -0.0002 | 0.0005 | 0.0008 | -0.0013 | -0.0025 | 0.0063 | 0.0100 |
| 42 | A' | 0.0797 | -0.0001 | -0.0004 | 0.0009 | -0.0010 | -0.0013 | -0.0050 | 0.0113 | -0.0125 |
| 43 | A' | 0.0808 | -0.0019 | 0.0012 | 0.0001 | -0.0005 | -0.0235 | 0.0149 | 0.0012 | -0.0062 |
| 44 | A' | 0.0831 | -0.0003 | -0.0002 | -0.0009 | 0.0008 | -0.0036 | -0.0024 | -0.0108 | 0.0096 |
| 45 | A" | 0.0880 | 0.0000 | 0.0000 | -0.0012 | -0.0059 | 0.0000 | 0.0000 | -0.0136 | -0.0670 |
| 46 | A' | 0.0882 | -0.0060 | -0.0150 | 0.0006 | -0.0005 | -0.0680 | -0.1701 | 0.0068 | -0.0057 |
| 47 | A" | 0.0885 | 0.0000 | 0.0000 | -0.0010 | 0.0008 | 0.0000 | 0.0000 | -0.0113 | 0.0090 |
| 48 | A" | 0.0885 | 0.0000 | 0.0000 | 0.0011 | -0.0049 | 0.0000 | 0.0000 | 0.0124 | -0.0554 |
| 49 | A" | 0.0923 | 0.0000 | 0.0000 | -0.0036 | 0.0009 | 0.0000 | 0.0000 | -0.0390 | 0.0098 |
| 50 | A' | 0.0948 | -0.0116 | 0.0078 | -0.0001 | -0.0007 | -0.1224 | 0.0823 | -0.0011 | -0.0074 |
| 51 | A" | 0.0973 | 0.0000 | 0.0000 | 0.0004 | 0.0000 | 0.0000 | 0.0000 | 0.0041 | 0.0000 |
| 52 | A" | 0.0973 | 0.0000 | 0.0000 | 0.0009 | -0.0014 | 0.0000 | 0.0000 | 0.0092 | -0.0144 |
| 53 | A" | 0.1021 | 0.0000 | 0.0000 | 0.0001 | -0.0044 | 0.0000 | 0.0000 | 0.0010 | -0.0431 |
| 54 | A' | 0.1035 | 0.0045 | -0.0017 | 0.0005 | 0.0002 | 0.0435 | -0.0164 | 0.0048 | 0.0019 |
| 55 | A" | 0.1072 | 0.0000 | 0.0000 | -0.0050 | 0.0016 | 0.0000 | 0.0000 | -0.0466 | 0.0149 |
| 56 | A" | 0.1076 | 0.0000 | 0.0000 | -0.0039 | 0.0012 | 0.0000 | 0.0000 | -0.0362 | 0.0112 |
| 57 | A" | 0.1081 | 0.0000 | 0.0000 | -0.0003 | 0.0018 | 0.0000 | 0.0000 | -0.0028 | 0.0167 |
| 58 | A" | 0.1082 | 0.0000 | 0.0000 | 0.0020 | -0.0042 | 0.0000 | 0.0000 | 0.0185 | -0.0388 |
| 59 | A' | 0.1088 | 0.0195 | -0.0011 | 0.0000 | 0.0007 | 0.1792 | -0.0101 | 0.0000 | 0.0064 |
| 60 | A" | 0.1156 | 0.0000 | 0.0000 | 0.0005 | -0.0019 | 0.0000 | 0.0000 | 0.0043 | -0.0164 |
| 61 | A" | 0.1169 | 0.0000 | 0.0000 | -0.0005 | -0.0084 | 0.0000 | 0.0000 | -0.0043 | -0.0719 |
| 62 | A" | 0.1176 | 0.0000 | 0.0000 | -0.0006 | 0.0016 | 0.0000 | 0.0000 | -0.0051 | 0.0136 |
| 63 | A" | 0.1176 | 0.0000 | 0.0000 | 0.0018 | -0.0048 | 0.0000 | 0.0000 | 0.0153 | -0.0408 |
| 64 | A' | 0.1199 | -0.0162 | 0.0042 | 0.0007 | -0.0007 | -0.1351 | 0.0350 | 0.0058 | -0.0058 |
| 65 | A" | 0.1228 | 0.0000 | 0.0000 | 0.0001 | 0.0017 | 0.0000 | 0.0000 | 0.0008 | 0.0138 |
| 66 | A" | 0.1229 | 0.0000 | 0.0000 | 0.0003 | 0.0016 | 0.0000 | 0.0000 | 0.0024 | 0.0130 |
| 67 | A" | 0.1232 | 0.0000 | 0.0000 | 0.0010 | 0.0015 | 0.0000 | 0.0000 | 0.0081 | 0.0122 |
| 68 | A" | 0.1232 | 0.0000 | 0.0000 | 0.0019 | -0.0005 | 0.0000 | 0.0000 | 0.0154 | -0.0041 |
| 69 | A" | 0.1247 | 0.0000 | 0.0000 | 0.0010 | -0.0019 | 0.0000 | 0.0000 | 0.0080 | -0.0152 |
| 70 | A" | 0.1265 | 0.0000 | 0.0000 | 0.0017 | -0.0017 | 0.0000 | 0.0000 | 0.0134 | -0.0134 |
| 71 | A" | 0.1265 | 0.0000 | 0.0000 | 0.0005 | 0.0014 | 0.0000 | 0.0000 | 0.0040 | 0.0111 |
| 72 | A' | 0.1271 | -0.0200 | -0.0440 | 0.0003 | -0.0011 | -0.1574 | -0.3462 | 0.0024 | -0.0087 |

| Mode $i$ | Symm | $\omega_i$/eV | $\kappa_i^{(1)}$/eV | $\kappa_i^{(2)}$/eV | $\gamma_i^{(1)}$/eV | $\gamma_i^{(2)}$/eV | $\kappa_i^{(1)}/\omega_i$ | $\kappa_i^{(2)}/\omega_i$ | $\gamma_i^{(1)}/\omega_i$ | $\gamma_i^{(2)}/\omega_i$ |
|---|---|---|---|---|---|---|---|---|---|---|
| 73 | A' | 0.1274 | 0.0191 | -0.0049 | 0.0003 | 0.0007 | 0.1499 | -0.0385 | 0.0024 | 0.0055 |
| 74 | A' | 0.1275 | 0.0064 | 0.0503 | 0.0010 | 0.0000 | 0.0502 | 0.3945 | 0.0078 | 0.0000 |
| 75 | A' | 0.1302 | 0.0015 | 0.0003 | 0.0013 | 0.0007 | 0.0115 | 0.0023 | 0.0100 | 0.0054 |
| 76 | A' | 0.1327 | -0.0066 | -0.0014 | 0.0011 | 0.0007 | -0.0497 | -0.0106 | 0.0083 | 0.0053 |
| 77 | A' | 0.1327 | 0.0018 | 0.0060 | 0.0010 | 0.0006 | 0.0136 | 0.0452 | 0.0075 | 0.0045 |
| 78 | A' | 0.1392 | 0.0002 | -0.0002 | 0.0012 | 0.0003 | 0.0014 | -0.0014 | 0.0086 | 0.0022 |
| 79 | A' | 0.1393 | 0.0000 | 0.0000 | 0.0009 | 0.0008 | 0.0000 | 0.0000 | 0.0065 | 0.0057 |
| 80 | A' | 0.1407 | 0.0037 | -0.0036 | 0.0010 | 0.0005 | 0.0263 | -0.0256 | 0.0071 | 0.0036 |
| 81 | A' | 0.1429 | -0.0011 | 0.0011 | 0.0007 | 0.0010 | -0.0077 | 0.0077 | 0.0049 | 0.0070 |
| 82 | A' | 0.1439 | 0.0709 | -0.0536 | 0.0020 | -0.0062 | 0.4927 | -0.3725 | 0.0139 | -0.0431 |
| 83 | A' | 0.1466 | -0.0905 | -0.0359 | -0.0020 | 0.0002 | -0.6173 | -0.2449 | -0.0136 | 0.0014 |
| 84 | A' | 0.1491 | -0.0464 | -0.0843 | -0.0026 | -0.0006 | -0.3112 | -0.5654 | -0.0174 | -0.0040 |
| 85 | A' | 0.1492 | -0.0013 | -0.0026 | 0.0014 | 0.0014 | -0.0087 | -0.0174 | 0.0094 | 0.0094 |
| 86 | A' | 0.1492 | 0.0001 | 0.0003 | 0.0016 | 0.0011 | 0.0007 | 0.0020 | 0.0107 | 0.0074 |
| 87 | A' | 0.1506 | -0.0123 | 0.0331 | 0.0010 | 0.0002 | -0.0817 | 0.2198 | 0.0066 | 0.0013 |
| 88 | A' | 0.1517 | -0.0053 | 0.0186 | 0.0007 | 0.0011 | -0.0349 | 0.1226 | 0.0046 | 0.0073 |
| 89 | A' | 0.1518 | 0.0167 | -0.0287 | 0.0010 | -0.0002 | 0.1100 | -0.1891 | 0.0066 | -0.0013 |
| 90 | A' | 0.1522 | -0.0452 | 0.0087 | 0.0004 | 0.0006 | -0.2970 | 0.0572 | 0.0026 | 0.0039 |
| 91 | A' | 0.1628 | -0.0092 | 0.0019 | 0.0032 | 0.0053 | -0.0565 | 0.0117 | 0.0197 | 0.0326 |
| 92 | A' | 0.1644 | 0.0016 | -0.0012 | 0.0197 | 0.0011 | 0.0097 | -0.0073 | 0.1198 | 0.0067 |
| 93 | A' | 0.1655 | 0.0204 | -0.0207 | 0.0019 | 0.0030 | 0.1233 | -0.1251 | 0.0115 | 0.0181 |
| 94 | A' | 0.1659 | 0.0000 | 0.0000 | 0.0058 | 0.0009 | 0.0000 | 0.0000 | 0.0350 | 0.0054 |
| 95 | A' | 0.1660 | 0.0021 | -0.0035 | 0.0014 | 0.0103 | 0.0127 | -0.0211 | 0.0084 | 0.0620 |
| 96 | A' | 0.1668 | 0.0005 | -0.0002 | 0.0006 | 0.0009 | 0.0030 | -0.0012 | 0.0036 | 0.0054 |
| 97 | A' | 0.1686 | 0.0230 | 0.0046 | -0.0009 | 0.0008 | 0.1364 | 0.0273 | -0.0053 | 0.0047 |
| 98 | A' | 0.1692 | -0.0098 | 0.0026 | 0.0028 | 0.0043 | -0.0579 | 0.0154 | 0.0165 | 0.0254 |
| 99 | A' | 0.1701 | -0.0001 | -0.0001 | 0.0023 | 0.0009 | -0.0006 | -0.0006 | 0.0135 | 0.0053 |
| 100 | A' | 0.1701 | 0.0004 | -0.0008 | 0.0013 | 0.0030 | 0.0024 | -0.0047 | 0.0076 | 0.0176 |
| 101 | A' | 0.1713 | -0.0111 | 0.0281 | -0.0008 | -0.0026 | -0.0648 | 0.1640 | -0.0047 | -0.0152 |
| 102 | A' | 0.1818 | -0.0004 | -0.0006 | 0.0008 | 0.0008 | -0.0022 | -0.0033 | 0.0044 | 0.0044 |
| 103 | A' | 0.1831 | -0.0070 | -0.0089 | -0.0004 | -0.0034 | -0.0382 | -0.0486 | -0.0022 | -0.0186 |
| 104 | A' | 0.1867 | 0.0001 | 0.0004 | 0.0015 | -0.0006 | 0.0005 | 0.0021 | 0.0080 | -0.0032 |
| 105 | A' | 0.1867 | 0.0000 | 0.0000 | 0.0004 | 0.0008 | 0.0000 | 0.0000 | 0.0021 | 0.0043 |

| Mode $i$ | Symm | $\omega_i$/eV | $\kappa_i^{(1)}$/eV | $\kappa_i^{(2)}$/eV | $\gamma_i^{(1)}$/eV | $\gamma_i^{(2)}$/eV | $\kappa_i^{(1)}/\omega_i$ | $\kappa_i^{(2)}/\omega_i$ | $\gamma_i^{(1)}/\omega_i$ | $\gamma_i^{(2)}/\omega_i$ |
|---|---|---|---|---|---|---|---|---|---|---|
| 106 | A' | 0.1915 | 0.0072 | -0.0370 | -0.0003 | -0.0015 | 0.0376 | -0.1932 | -0.0016 | -0.0078 |
| 107 | A' | 0.1931 | 0.0220 | -0.0102 | -0.0012 | 0.0006 | 0.1139 | -0.0528 | -0.0062 | 0.0031 |
| 108 | A' | 0.1940 | -0.0051 | -0.0355 | 0.0008 | -0.0024 | -0.0263 | -0.1830 | 0.0041 | -0.0124 |
| 109 | A' | 0.1963 | -0.0089 | -0.0005 | -0.0028 | 0.0004 | -0.0453 | -0.0025 | -0.0143 | 0.0020 |
| 110 | A' | 0.2018 | -0.0002 | -0.0026 | -0.0059 | 0.0007 | -0.0010 | -0.0129 | -0.0292 | 0.0035 |
| 111 | A' | 0.2059 | -0.0387 | -0.0895 | -0.0010 | -0.0090 | -0.1880 | -0.4347 | -0.0049 | -0.0437 |
| 112 | A' | 0.2064 | 0.0000 | 0.0002 | -0.0003 | 0.0008 | 0.0000 | 0.0010 | -0.0015 | 0.0039 |
| 113 | A' | 0.2065 | -0.0041 | -0.0140 | 0.0010 | -0.0055 | -0.0199 | -0.0678 | 0.0048 | -0.0266 |
| 114 | A' | 0.2092 | 0.0923 | -0.0860 | -0.0015 | -0.0185 | 0.4412 | -0.4111 | -0.0072 | -0.0884 |
| 115 | A' | 0.2102 | 0.1085 | 0.0285 | -0.0039 | 0.0001 | 0.5162 | 0.1356 | -0.0186 | 0.0005 |
| 116 | A' | 0.2106 | 0.0173 | 0.0657 | -0.0001 | -0.0103 | 0.0821 | 0.3120 | -0.0005 | -0.0489 |
| 117 | A' | 0.2113 | -0.1168 | -0.0092 | -0.0058 | -0.0001 | -0.5528 | -0.0435 | -0.0274 | -0.0005 |
| 118 | A' | 0.2935 | 0.2175 | -0.0614 | 0.0152 | 0.0699 | 0.7411 | -0.2092 | 0.0518 | 0.2382 |
| 119 | A' | 0.2943 | 0.0349 | -0.0360 | -0.0419 | -0.0023 | 0.1186 | -0.1223 | -0.1424 | -0.0078 |
| 120 | A' | 0.2945 | -0.0827 | -0.3230 | 0.0203 | -0.0197 | -0.2808 | -1.0968 | 0.0689 | -0.0669 |
| 121 | A' | 0.3984 | 0.0012 | -0.0032 | 0.0020 | 0.0018 | 0.0030 | -0.0080 | 0.0050 | 0.0045 |
| 122 | A' | 0.3984 | -0.0041 | -0.0020 | 0.0022 | 0.0020 | -0.0103 | -0.0050 | 0.0055 | 0.0050 |
| 123 | A' | 0.3994 | 0.0002 | 0.0000 | 0.0021 | 0.0018 | 0.0005 | 0.0000 | 0.0053 | 0.0045 |
| 124 | A' | 0.3995 | -0.0137 | -0.0174 | 0.0029 | 0.0023 | -0.0343 | -0.0436 | 0.0073 | 0.0058 |
| 125 | A' | 0.3996 | -0.0001 | -0.0004 | 0.0019 | 0.0014 | -0.0003 | -0.0010 | 0.0048 | 0.0035 |
| 126 | A' | 0.4003 | -0.0009 | -0.0009 | 0.0019 | 0.0016 | -0.0022 | -0.0022 | 0.0047 | 0.0040 |
| 127 | A' | 0.4004 | 0.0018 | 0.0021 | 0.0018 | 0.0016 | 0.0045 | 0.0052 | 0.0045 | 0.0040 |
| 128 | A' | 0.4007 | 0.0015 | 0.0030 | 0.0019 | 0.0015 | 0.0037 | 0.0075 | 0.0047 | 0.0037 |
| 129 | A' | 0.4007 | -0.0010 | 0.0000 | 0.0020 | 0.0019 | -0.0025 | 0.0000 | 0.0050 | 0.0047 |
| 130 | A' | 0.4015 | -0.0037 | -0.0038 | 0.0029 | 0.0021 | -0.0092 | -0.0095 | 0.0072 | 0.0052 |
| 131 | A' | 0.4015 | -0.0017 | -0.0017 | 0.0020 | 0.0018 | -0.0042 | -0.0042 | 0.0050 | 0.0045 |
| 132 | A' | 0.4017 | 0.0004 | 0.0003 | 0.0019 | 0.0014 | 0.0010 | 0.0007 | 0.0047 | 0.0035 |
| 133 | A' | 0.4023 | -0.0108 | -0.0130 | 0.0016 | 0.0014 | -0.0268 | -0.0323 | 0.0040 | 0.0035 |
| 134 | A' | 0.4023 | -0.0014 | -0.0018 | 0.0017 | 0.0014 | -0.0035 | -0.0045 | 0.0042 | 0.0035 |
| 135 | A' | 0.4023 | 0.0130 | 0.0068 | 0.0015 | 0.0011 | 0.0323 | 0.0169 | 0.0037 | 0.0027 |
| 136 | A' | 0.4024 | 0.0094 | 0.0118 | 0.0025 | 0.0019 | 0.0234 | 0.0293 | 0.0062 | 0.0047 |
| 137 | A' | 0.4025 | 0.0086 | 0.0150 | 0.0017 | 0.0015 | 0.0214 | 0.0373 | 0.0042 | 0.0037 |
| 138 | A' | 0.4026 | 0.0111 | 0.0077 | 0.0018 | 0.0013 | 0.0276 | 0.0191 | 0.0045 | 0.0032 |